\documentclass[a4paper,11pt]{article}
\usepackage{jheppub} 
\usepackage{lineno}
 \usepackage{graphics}
 \usepackage{float}
 \usepackage{tikz-cd} 
\usepackage{url}
\usepackage{hyperref}
\usepackage{float}
\usepackage{standalone}
\usepackage{bm}
\usepackage{quiver}
\usepackage{cleveref}
\usepackage{subcaption}
\usepackage{graphicx} 
\usepackage{array}
\usepackage{paralist}
\usepackage{amsmath}
\usepackage{physics}
\usepackage{float}
\usepackage{xcolor}
\usepackage{amsfonts}
\usepackage{amssymb}
\usepackage{verbatim}
 \usepackage{indentfirst}
\usepackage{tikz}
\usepackage{soul}
\usepackage{pifont}
\usepackage{forest}
\usepackage{tikz-cd}
\tikzcdset{
	arrow style=tikz,
	diagrams={>={Straight Barb[scale=0.6]}},
    every arrow/.append style={line width=0.4pt}
}
\usetikzlibrary{calc}
\tikzset{curve/.style={settings={#1},to path={(\tikztostart)
    .. controls ($(\tikztostart)!\pv{pos}!(\tikztotarget)!\pv{height}!270:(\tikztotarget)$)
    and ($(\tikztostart)!1-\pv{pos}!(\tikztotarget)!\pv{height}!270:(\tikztotarget)$)
    .. (\tikztotarget)\tikztonodes}},
    settings/.code={\tikzset{quiver/.cd,#1}
        \def\pv##1{\pgfkeysvalueof{/tikz/quiver/##1}}},
    quiver/.cd,pos/.initial=0.25,height/.initial=0}
\tikzset{
	solid node/.style={circle,draw,inner sep=1.2,fill=black},
	hollow node/.style={circle,draw,inner sep=1.2},
	left label/.style={above left,midway},
	right label/.style={above right,midway}
}
\usepackage{caption, subcaption}
 
 \usepackage{tikz}
\usepackage{multirow} 
\usepackage{makecell}
\usepackage{longtable}

\def\beq{\begin{eqnarray}}
\def\eeq{\end{eqnarray}}
\def\beq#1\eeq{\begin{align}#1\end{align}}

\title{On duality of four dimensional $\mathcal{N}=1$ gauge theory }

\author[]{Yuanyuan Fang, Jing Feng, Dan Xie}
\affiliation[]{Department of Mathematics, Tsinghua University, Beijing, 100084, China}

\abstract{We show that Seiberg-like duality of $\mathcal{N}=1$ gauge theory coupled with tensor chiral fields and fundamental chiral fields works if the meson spectrum built from 
the tensor fields takes particular form: a) It should be truncated; b) The $R$ charges of tensor fields $\{R_a\}$ and the truncated mesons $\{R_j\}$ take  very special values. The meson spectrum so that the duality works is encoded elegantly in the factorization of the polynomial $y^n-1=\Phi_{+}\Phi_{-}$. Our consideration covers many known $\mathcal{N}=1$ dualities and 
generates a large class of new examples.}

\begin{document} 
\maketitle
\flushbottom

\section{Introduction}
One of the most important discoveries in the study of 4d $\mathcal{N}=1$ non-abelian gauge theory was the electric-magnetic 
duality of SQCD found by Seiberg \cite{Seiberg:1994pq}. The basic features of the Seiberg duality are the following: a) The gauge group 
is different in the dual description; b)
The composite chiral operator of one theory is mapped to the elementary field (gauge singlet) of the dual theory; c) A superpotential is needed so that the composite chiral operators of the dual theory are projected out.
Seiberg duality plays a crucial role in understanding the strong coupling dynamics of non-abelian gauge theory \cite{Intriligator:1995au}. 

Seiberg duality was soon generalized to many other models: theory with one adjoint chiral superfield \cite{Kutasov:1995ve, Kutasov:1995np}, and theory with two adjoint chiral superfields \cite{Brodie:1996vx}, and many other similar models in \cite{Intriligator:1995ax, Intriligator:1995id,Pouliot:1995me,Brodie:1996xm,Pouliot:1995sk,Pouliot:1996zh}. The basic picture of the duality is the same as the Seiberg duality, for example, the composite chiral operators are mapped to the elementary fields, and often a dual superpotential is needed. 

A crucial ingredient in all the above models is the following: there is a truncation for the chiral spectrum of the mesons built from the tensor field (such as the adjoint chiral of. The truncation was often introduced by a superpotential for the tensor fields, for example, the $A_k$ or ($D_{k+1},~k$ odd) type superpotential for the adjoint chirals of $\mathcal{N}=1$ theory. However, one already needs to impose the so-called quantum constraints to truncate the meson spectrum. Such a strategy has been used in \cite{Kutasov:2014yqa} to find the dual description of the so-called $E_7$ model \cite{Intriligator:2003mi} for adjoint SQCD.

The purpose of this paper is to explore the following question: for what kind of truncation of meson spectrum
one can find a sensible dual theory. Our main discovery is that the duality is essentially controlled by the truncation of the meson spectrum built from tensorial matter. Assuming our model is given by $SU(N_c)$ gauge group coupled with $N_f$ flavors of fundamental matter, and $N_A$ adjoint matters. The basic picture of the duality is shown in Figure \ref{basicdual}. The duality works if the $R$ charges of the adjoint matter and the truncated set of mesons satisfy the equation \footnote{This equation has been discussed in \cite{Kutasov:2014wwa,Bajc:2019vbp}, and is derived from the match of large $N_c, N_f$ limit of the superconformal index of dual theory. Our point here is that this equation will ensure that the duality works for the finite $N_c, N_f$ limit.}
\begin{equation}
\sum_{j=1}^\alpha t^{R_j}=\frac{t^{\Delta+2}-1}{-1+t^2+\sum_a(t^{R_a}-t^{2-R_a})}\,.
\label{factorization}
\end{equation}
Here $\Delta$ is the pairing constant of mesons built from adjoint chirals, and $R_a$'s are the charges for the adjoint chirals, and $R_j$'s are the charges for the mesons.  
Once the $R$ charges satisfy the above equation, we can check that various physical quantifies for dual theory also match: a) the anomalies of the $R$ symmetry such as $\Tr(R)$ and $\Tr(R^3)$ agree, which implies the central charges $a,c$ agree; b) the spectrums of mesonic and baryonic operators are mapped perfectly. 

It is then quite amazing that as long as the $R$ charges of the adjoint chirals $R_a$ and the truncated spectrum $R_j$ are derived from the factorization of the polynomial $y^n-1=\Phi^{+}\Phi^{-}$ (where $\Phi^{+}$ is a polynomial with positive coefficients), one can get an $\mathcal{N}=1$ duality. So one immediately gets a large class of new types of $\mathcal{N}=1$ dualities. Moreover, for each factorization, one can get many new dualities by distributing the gauge singlets in electric and magnetic theory differently.

For other classical gauge groups with tensor fields, one often needs to introduce an involution 
operation on the above set of $R$ charges so that duality could work. We also studied 
the duality involving exceptional gauge groups. The generalization to semi-simple groups such as the quiver 
gauge theory is straightforward as one can perform the duality on each gauge group separately. 

The present paper mainly focuses on the combinatorial nature of $\mathcal{N}=1$ duality. The dynamical questions such as whether such truncation is possible or not would be left for future exploration. 

This paper is organized as follows: Section \ref{generaldis} discusses the basic feature of $\mathcal{N}=1$ duality of 
$SU(N_c)$ gauge theory coupled with adjoint chirals and $N_f$ chirals; Section \ref{simplegaugegp} discusses the generalization to 
simple gauge groups; Section \ref{semisimple} discusses the duality for theory with semi-simple gauge groups, such as quiver gauge theory; A conclusion is given in Section \ref{conclusion}.

\section{General discussion}\label{generaldis}

\subsection{Basic picture of $\mathcal{N}=1$ duality}
\label{general}
\begin{figure}[H]
	\centering
       \begin{tikzcd}
       |[draw,rectangle]| N_f\ar[r,bend left=10,"c"]&    
       |[draw,circle]| N_c \ar[l,bend left=10,"d"]\ar[loop,out =290,in=250,looseness=5,"v",no head]\ar[loop,out =110,in=70,looseness=5,"u",no head]
       \end{tikzcd}
       \begin{tikzcd}
		\stackrel{}{\Longrightarrow}    
	\end{tikzcd}       
    \begin{tikzcd}
       |[draw,rectangle]| N_f\ar[r,bend right=10,"d^*"']\ar[loop,out =290,in=250,looseness=9,"",no head]\ar[loop,out =140,in=170,looseness=9,"",no head]\ar[loop,out =200,in=230,looseness=9,"",no head]\ar[loop,out =110,in=70,looseness=9,"{[M_I]}",no head]& 
       |[draw,circle]|  N_c^{'} \ar[l,bend right=10,"c^*"']\ar[loop,out =290,in=250,looseness=5,"v^*",no head]\ar[loop,out =110,in=70,looseness=5,"u^*",no head]
       \end{tikzcd}
       \caption{Basic picture of $\mathcal{N}=1$ duality.}
        \label{basicdual}
\end{figure}
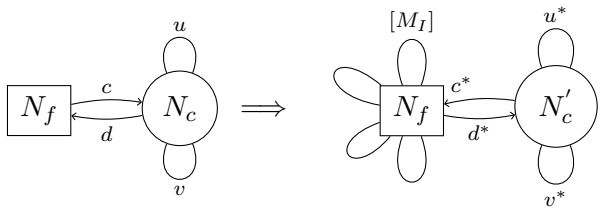
Let's consider duality for  $SU(N_c)$ gauge theory coupled with two adjoint chiral fields $u,v$ \footnote{Here we use two adjoint chirals to simplify the illustration, and all the following discussions are still valid for any number of adjoint fields.} and $N_f$ pairs of fundamental and anti-fundamental chiral fields $c,d$.  Notice that this gauge theory might be a part of a  quiver gauge theory. There might be marginal superpotential $W_E$ for the electric model.
The duality works as follows:
\begin{enumerate}
\item  Replace every chiral field of the original theory with a dual chiral field: $c\to c^*$, $d\to d^*$, $u\to u^*$, $v\to v^*$. And the dual fields are in the conjugate representation of gauge group: the orientation of the fundamental and anti-fundamental field is reversed, see Figure \ref{basicdual}.
\item For each oriented loop starting with the flavor node, there is an associated dressed mesonic chiral field, i.e. $M_I=cu^{a_1}v^{b_1} u^{a_2} v^{b_2}\ldots d$. One adds a dual gauge singlet $[M_I]$ which is now in adjoint representation of the flavor node. Here $I$ is a sequence of integers $[a_1, b_1, a_2, b_2,\ldots]$.
\item The superpotential term is changed as follows: replace the combination of letters $M_I$ in $W_E$ by the gauge singlet $[M_I]$, and add new terms in the superpotential as
\begin{equation}
\Delta W=\sum_I [M_I] d^*u^{a_1^{'}}v^{b_1^{'}} \ldots c^*=\sum [M_I]M_{I^{'}}\,.
\label{dualsuper}
\end{equation}
Here $M_{I^{'}}$ is the composite dressed meson in the dual theory. 
\item Finally, one might be able to integrate out massive fields using the superpotential in the magnetic theory.
\end{enumerate}
To make the above duality proposal work, the set of undressed mesons (which is formed by just adjoint chiral fields) should satisfy the following constraints: a)  
The dual theory has a finite number of gauge singlets, which means the set of letters involving $u,v$ 
should be finite, i.e. there are truncations on the set of mesons. The superpotential $W_E$ is useful in 
truncating the set of mesons. However, it is often necessary to use quantum constraints to truncate the mesons. b) The $R$ charges of the allowed mesons $U_I=u^{a_1}v^{b_1} u^{a_2} v^{b_2}\ldots$ are paired: 
\begin{equation}
R(U_I)+R(U_{I^{'}})=\Delta\,.
\label{pair}
\end{equation}
Here $\Delta$ is a fixed constant. This condition is required so that the added superpotential (\ref{dualsuper})
is marginal. This superpotential is crucial as 
the equation of motion for the gauge singlet $[M_{I}]$ would set the composite $M_{I^{'}}$ operator of the dual theory to be zero in the chiral ring. The duality then maps the composite chiral operator to an elementary 
chiral operator in the dual theory, and there are no composite chiral operators in the dual theory.

The rank $N_c^{'}$ of the dual gauge theory can be computed easily from the above general setup of 
the duality. First, let's fix the $R$ charges of adjoint chiral fields $u$ and $v$ as $[u],[v]$ (it might be fixed by a superpotential term $f(u,v)$ in $W_E$). 
The anomaly free condition for the $U(1)_R$ symmetry of electric theory is 
\begin{align*}
&([u]-1)+([v]-1)+\frac{N_f}{N_c}(R_c-1)+1=0, \nonumber\\
\Rightarrow~&R_c=1-([u]+[v]-1)\frac{N_c}{N_f}\,.
\end{align*}
Now for the dual theory, one has the new superpotential term (\ref{dualsuper}), which should also have $R$ charge 2:
\begin{equation*}
2R_c+R(U_I)+R(U_{I^{'}})+2R_{c^*}=2\,.
\end{equation*}
Using the pairing condition for the mesons, the $R$ charge of the $c^*$ field is:
\begin{equation}
R_{c^*}=1-R_c-\frac{\Delta}{2}\,.
\label{fundarcharge}
\end{equation}
This charge might be negative, but it would not cause the problem as long as the gauge invariant operator 
has positive $R$ charge. We can now compute the rank $N_c^{'}$ of the dual gauge theory by requiring the anomaly free condition of the dual $U(1)_R$ symmetry:
\begin{align*}
& ([u]+[v]-1)+\frac{N_f}{N_c^{'}}(R_{c^*}-1)=0, \nonumber\\
\Rightarrow~&N_c^{'}=\frac{N_f(R_c+\frac{\Delta}{2})}{[u]+[v]-1}\nonumber\\
&=\frac{N_f-N_c([u]+[v]-1)+N_f\frac{\Delta}{2}}{[u]+[v]-1}\nonumber\\
&=N_f\frac{2+\Delta}{2([u]+[v]-1)}-N_c=aN_f-N_c\,.
\end{align*}
The crucial number $a$ is determined by the $U(1)_R$ charge of $u,v$ and the pairing constant $\Delta$ of the meson spectrum:
\begin{equation}\label{aequation}
\boxed{a=\frac{2+\Delta}{2([u]+[v]-1)}\,.}
\end{equation}
The above formula makes sense if 
\begin{equation}
[u]+[v]>1\,.
\label{uvrcharge}
\end{equation}

The chiral spectrum is matched as follows: First, the original meson $M_I$ is mapped to the gauge singlet $[M_I]$, and the extra meson $M_{I^{'}} $ in the dual theory 
is projected out by using the added new superpotential term. The scaling dimensions of 
these chiral operators are mapped straightforwardly. 

The baryon spectrum is mapped as follows \cite{Kutasov:2014yqa}: One first has dressed quark $c_I=U_Ic$, and 
the baryons are formed as $B^{l_1,\ldots, l_a}=c_1^{l_1}\ldots c^{l_a}_a$ with $\sum_{i=1}^a l_i=N_c$.
They are simply built as the determinant of the matrix formed by dressed quarks. The total number of baryons is $C_{aN_f}^{N_c}$.

The dual baryon is formed by dual dressed quark $c_I^*=U_Ic^*$, and $B^{\tilde{l}_1,\ldots, \tilde{l}_a}=c_1^{*\tilde{l}_1}\ldots c^{*\tilde{l}_a}_a$, with  $\sum_{i=1}^a \tilde{l}_i=N_c^{'}$. The number $\tilde{l}_i=N_f-l_{I^d}$, with $I^d$ the paired meson for $I$. 

Let's now verify that their $R$ charges are the same (so the scaling dimension is also the same). The $R$ charge 
for electric baryon $B^{l_1,\ldots, l_a}$ is 
\begin{equation*}
\sum l_j R_j+N_c R_c\,.
\end{equation*}
The R charge for dual baryon $B^{\tilde{l}_1,\ldots, \tilde{l}_a}$ is 
\begin{align*}
&\sum \tilde{l}_j R_j+N_c^{'} R_{c^{*}}=\sum (N_f-l_{j^d}) (\Delta-R_{j^d})+(aN_f-N_c)(1-R_c-{\Delta\over 2}) \nonumber\\
&=N_cR_c+\sum l_j R_j\,.
\end{align*}
Firstly we used $N_c^{'}=aN_f-N_c,~~R_{c^{*}}=1-R_c-{\Delta\over 2}$. Next 
we used $\sum_j R_j=\frac{a \Delta}{2}, R_c=1-\frac{x N_c} {N_f}$ and $a=\frac{\Delta+2}{2x}$, with $x=\sum_u(R_u-1)+1$ to verify the equality of the $R$ charges of the baryons.

\subsection{Truncation by superpotential}
We have seen in the last subsection that $\mathcal{N}=1$ duality might work if 
the set of mesons involving adjoint chiral fields takes a particular form of truncation. 
We'd like to find all the consistent truncations satisfying the condition (\ref{pair}). 

The truncation can be achieved by using a superpotential $f(u,v)$. The superpotential is assumed to be a trace on two matrices $u,v$. We take an algebraic approach by treating 
$u,v$ as two non-commutative variables. Each term in the potential $f(u,v)=\sum_\alpha \phi_\alpha(u,v)$ has a cyclic equivalence (remember that the trace function on the matrices has this property). The derivative of the superpotential is defined as 
\begin{equation}
f_u=\sum_\alpha {\partial_u}\phi_\alpha(u,v)=\sum_\alpha[\partial_u(\sum_{cyc} u\ldots )]=\sum_\alpha\sum_{cyc} \ldots\,.
\label{cyc}
\end{equation}
Here the sum is over the cyclic equivalence classes of $\phi_\alpha$ such that the first term is $u$. We then have a non-commutative algebra
\begin{equation*}
J_f={C[u,v]\over \{f_u,f_v\}},
\end{equation*}
We are looking for $f$ such that $J_f$ is finite dimensional, and $f$ is required to be a quasi-homogeneous polynomial (the $R$ charge of $f$ is two) so that there is grading for field $u$ and $v$. 
 
A necessary condition for the classical truncation is that $f(u,v)$ should define an isolated singularity if one regards $u, v$ as commutative variables, so that $J_f$ is finite dimensional if it is regarded as a commutative Jacobian algebra. The constraint (\ref{uvrcharge}) would mean that  
 $f(u,v)$ defines an isolated ADE singularity  \footnote{Usually the ADE singularity is defined as surface singularity $z^2+f(x,y)=0$ and the ADE condition implies $[z]+[x]+[y]>d$ assuming the weight of the polynomial is $d$. In our case, the weight of $f$ is two, and the weight of $[z]$ is one, and so the ADE condition implies the weights of $[x]$ and $[y]$ satisfy $[x]+[y]>2-[z]=1$.}. 

However, not every ADE singularity gives rise to a good truncation if $u, v$ are regarded as non-commutative 
variables. For example, the $E_6$ potential $f=u^3+v^4$ would give the relation
\begin{equation*}
u^2=0,~~v^3=0.
\end{equation*}
As a commutative algebra, $J_{E_6}$ is finite dimensional and has only 6 elements which are given by $\{1, u, v, v^2, uv, uv^2\}$. However, as a non-commutative algebra, the $J_f$ has an infinite number of elements, i.e words like $uvuvuvuv\ldots$ are nonzero simply because $u, v$ are noncommutative. 

The set of ADE singularities that do give consistent truncation is listed in Table \ref{potential}. Notice that 
for $E_7$, $D_k(k\text{ even})$, one needs to impose quantum constraints to get consistent truncation. One 
cannot get consistent truncation for $E_6, E_8$ superpotential.
 
\begin{table}[H]
    \centering
    \begin{tabular}{|c|c|c|c|c|}
    \hline
Duality&$f(u,v)$    & $([u],[v])$ & $\Delta$ & $N_c^{\prime}$\\
  \hline
  $A_1$ & $u^2+v^2$   & (1,1)&0&$N_c^{\prime}=N_f-N_c$\\
 \hline
 $A_k$ &$u^2+v^{k+1}$   & $(1,\frac{2}{k+1})$&$\frac{2k-2}{k+1}$&$N_c^{\prime}=k N_f-N_c$\\
 \hline
  \textcolor{red}{$D_k$}  &$u^{k+1}+uv^2$   & $(\frac{2}{k+1},\frac{k}{k+1})$&$\frac{4k-2}{k+1}$&$N_c^{\prime}=3k N_f-N_c$\\
  \hline
 \textcolor{red}{$E_7$ } &$u^3+uv^3$  & $(\frac{2}{3},\frac{4}{9})$&$\frac{14}{3}$&$N_c^{\prime}=30 N_f-N_c$\\
  \hline
    \end{tabular}
    \caption{Summary of superpotential $f(u,v)$ which would give good truncation of mesons. Quantum constraints need to be imposed for $D_k(k\text{ even})$ and $E_7$. }
    \label{potential}
\end{table}

\textbf{Example}: Consider $D_{k}$ superpotential $f(u,v)=u^{k+1}+uv^2$, and the ideal of $J_f$ is given (see equation (\ref{cyc})):
\begin{align}
&\frac{\partial f}{\partial v}=uv+vu=0,  \nonumber\\
&\frac{\partial f}{\partial u}=(k+1)u^{k}+v^2=0.
\label{Dconstr}
\end{align}
The first equation simply means that $u$ and $v$ are anticommuting, and the second equation implies that $u^{k}$ 
is the same as $v^2$ in the chiral ring. The following set is certainly the part of  Jacobian algebra $J_f$ 
\begin{align}
&\{1,u,\ldots, u^{k-1}\}, \nonumber\\
& \{v, vu, \ldots, vu^{k-1}\}, \nonumber\\
&\{v^2,v^2u,\ldots, v^2u^{k-1}\}. \nonumber \\
\label{Dtype}
\end{align}
Let's now consider $v^3$, one gets from two equations in (\ref{Dconstr}):
\begin{equation*}
(k+1)(u^{k}v+vu^k)+2v^3=0 \to v^3=-\frac{k+1}{2}(u^kv+(-1)^ku^kv)\,.
\end{equation*}
So for $k$ odd, we have $v^3=0$, and  the algebra $J_f$ has $3k$ elements, see (\ref{Dtype}). For $k$ even, one needs to truncate the set by assuming quantum constraint $v^3=0$. It is easy to check that the mesons have 
the desired pairing and the paring constant is listed in Table \ref{potential}.

\subsection{Truncation from equality of superconformal index}
Instead of using superpotential to find the consistent truncation of the meson spectrum, one can find 
a large class of new examples by using superconformal index \cite{Kutasov:2014wwa,Bajc:2019vbp}. 
See \cite{dolan2009applications,Spiridonov:2009za} for more details on the superconformal index.

In the large $N_c, N_f$ limit, the equality of the index for two dual gauge theories in Figure \ref{basicdual} takes the form
\begin{equation*}
{g_E\bar{g}_E-g_{M}\bar{g}_M \over (1-f)}=h_M-h_E\,.
\end{equation*}
Here $f$ is the contribution of the vector multiplet and adjoint chiral, $(g_E,g_{M})$ are the contribution of 
the fundamental chirals, and $(\bar{g}_E, \bar{g}_M)$ are the contribution of the anti-fundamental chirals,
$(h_E, h_M)$ are the contribution of gauge singlets. The detailed formulas for various fields are given in the appendix \ref{index}. Here $E$ denotes the contribution in the electric frame, and $M$ denotes the contribution in the magnetic frame.
The equality of the index gives the equation:
\begin{align}
&(t^{R_Q}-t^{2-R_Q})^2-(t^{R_q}-t^{2-R_q})^2=(1-t^2-\sum_a(t^{R_a}-t^{2-R_a}))(\sum_j t^{2R_Q+R_j}-t^{2-(2R_Q+R_j)}), \nonumber\\
\Rightarrow~&\frac{t^{2R_Q}+t^{4-2R_Q}-t^{2R_q}-t^{4-2R_q}}{(1-t^2-\sum_a(t^{R_a}-t^{2-R_a}))}=\sum_j (t^{2R_Q+R_j}-t^{2-(2R_Q+R_j)})\,.
\label{centralequation}
\end{align}

Here $R_Q$ ($R_q$) is the $R$ charge for the fundamental fields in an electric (magnetic) frame. If there is a pairing between the mesons $R_j+R_j^{'}=\Delta$, 
 then we have $\sum_j t^{R_j}= t^\Delta\sum_j t^{-R_j}$, and $R_Q+R_q=1-\frac{\Delta}{2}$ (see equation (\ref{fundarcharge})). We then have the equation
\begin{equation*}
\frac{(t^{2R_Q}-t^{2R_q})-t^{\Delta+2}(t^{2R_Q}-t^{2R_q})}{(1-t^2-\sum_a(t^{R_a}-t^{2-R_a}))}=[t^{2R_Q}-t^{2R_q}]\sum_j t^{R_j}\,.
\end{equation*}

So the equality of the index is satisfied if we have 
\begin{equation}\label{ind}
\boxed{\sum_{j=1}^\alpha t^{R_j}=\frac{t^{\Delta+2}-1}{-1+t^2+\sum_a(t^{R_a}-t^{2-R_a})}\,.}
\end{equation}

We see that whether duality works depends on the factorization of the polynomial 
\begin{equation*}
y^n-1=\prod_{d|n}\Phi_d(y)\,.
\end{equation*}
Here $\Phi_d(y)$ is the cyclotomic polynomial. To find consistent truncation of meson, one just needs to factorize the cyclotomic polynomial into a positive part (all the coefficients are positive) and the negative part \cite{Bajc:2019vbp}: 
\begin{equation*}
y^n-1=\Phi_{+}\Phi_{-}\,.
\end{equation*}
From the factorization, one can find the data $(\Delta, R_a, R_j)$ as follows. First, assuming the highest order of $\Phi_{-}$ is $n_1$ which is paired with the constant term $-1$, we can find the normalization constant $\delta={2\over n_1}$. Then we have
\begin{enumerate}
\item The pairing constant is $\Delta=n\delta-2=\frac{2}{n_1}(n-n_1)$.
\item The $R$ charges $R_a$ for the adjoint fields $u_a$ are found from the positive exponents $(c_1, c_2, \ldots)$  of $\Phi_{-}$ as $(c_1\delta, c_2\delta,\ldots)$.
\item The $R$ charges  $R_j$ for the mesons built from adjoint fields are found from the exponents $(d_0, d_1,\ldots,)$ of $\Phi_{+}$ as $(d_0 \delta, d_1 \delta,\ldots )$.

\end{enumerate}
This is the complete set of data needed for establishing $\mathcal{N}=1$ duality. Here are some remarks:
\begin{enumerate}
\item Although the equality of the index is computed in the large $N_c, N_f$ limit, one can check the central charges agree for the finite $N_c, N_f$ value, see section \ref{generalcentral}.
\item Once we find the $R$ charges for the adjoint fields, it might be possible to add the marginal term for the adjoint fields $u_a$. One certainly recovers the known ADE superpotential as a small subset. However, one should not expect to get the mesonic spectrum from the classical relation of the superpotential. Rather, one may simply 
impose the quantum constraint by hand.
\item The construction is symmetric under the exchange of electric and magnetic form, and so the duality is an involution. We will discuss more about this in the next subsection.
\item Notice that here we only get the $R$ charges of the mesons, and knowledge about how they are formed from the adjoint chiral fields $u_a$ is not clear: a) there might be more than one possible combinations of a given $R$ charge $R_j$; b) even if the combination of the $u_a$ field is given, the ordering of the fields $u_a$ is not clear as they are non-commutative variables. 
\end{enumerate}
All possible factorizations of $y^8-1$ are shown in Table \ref{cyclotomic8}.

\begin{table}[H]
    \centering
    \resizebox{\linewidth}{!}{
    \begin{tabular}{|c|c|c|}
    \hline
$\Phi_8^+ (y)$&$\Phi_8^- (y)$ & $R_a, R_j$  \\
  \hline
$\Phi_2(y)=y+1$& $\Phi_1(y)\Phi_4(y)\Phi_8(y)=y^7-y^6+y^5-y^4+y^3-y^2+y-1$ & $(\frac{2}{7},\frac{6}{7},\frac{10}{7}),(\frac{2}{7},0)$   \\
$\Phi_4(y)=y^2+1$& $\Phi_1(y)\Phi_2(y)\Phi_8(y)=y^6-y^4+y^2-1$ & $(\frac{2}{3}),(\frac{2}{3},0)$ \\
$\Phi_8(y)=y^4+1$& $\Phi_1(y)\Phi_2(y)\Phi_4(y)=y^4-1$ & $(1),(2,0)$ \\
$\Phi_2(y)\Phi_4(y)=y^3+y^2+y+1$& $\Phi_1(y)\Phi_8(y)=y^5-y^4+y-1$ & $(\frac{2}{5}),(\frac{3}{5},\frac{2}{5},\frac{1}{5},0)$ \\
$\Phi_2(y)\Phi_8(y)=y^5+y^4+y+1$& $\Phi_1(y)\Phi_4(y)=y^3-y^2+y-1$ & $(\frac{2}{3}),(\frac{10}{3},\frac{8}{3},\frac{2}{3},0)$ \\
$\Phi_4(y)\Phi_8(y)=y^6+y^4+y^2+1$& $\Phi_1(y)\Phi_2(y)=y^2-1$ & $(1),(6,4,2,0)$ \\
$\Phi_2(y)\Phi_4(y)\Phi_8(y)=y^7+y^6+y^5+y^4+y^3+y^2+y+1$& $\Phi_1(y)=y-1$ & $(1),(14,12,10,8,6,4,2,0)$ \\
\hline
    \end{tabular}}
    \caption{The possible products of cyclotomic polynomials with all coefficients positive
$\Phi_n^+ (y)$ and the corresponding antipalindromic polynomials $\Phi_n^-(y)=(y^n-1)/\Phi_n^+(y)$ for $n = 8$. }
    \label{cyclotomic8}
\end{table}

\textbf{Remark}: For the given ADE $R_a$ charges of the adjoint fields listed in Table \ref{potential}, it is possible to find an infinite sequence of $R_j$ charges and the paring constant $\Delta_n^{ADE}$. These may be thought of as the rank $n$ version of ADE theory. For example, if $R(u)=\frac{2}{3}$, then the right hand side of equation (\ref{ind}) becomes ${t^{\Delta+2}-1\over -1+t^2+t^{2/3}-t^{4/3}}={y^{\frac{3\Delta}{2}+3}-1\over -1+y^3+y-y^2}={y^{\frac{3\Delta}{2}+3}-1\over \Phi_1 \Phi_4}$, so as long as $\frac{3\Delta}{2}+3$ has a divisor $4$, one can have a consistent factorization. It would be interesting to realize these spectrums from some physical models.

\subsection{More dualities}
More generally, one may assume that there are already gauge singlets in the electric frame, and the right hand side of the equation (\ref{centralequation}) becomes 
\begin{align}
\sum_j( t^{2R_Q+R_j^M}-t^{2-(2R_Q+R_j^M)})-\sum_l(t^{2R_q+R_l^E}-t^{2-(2R_q+R_l^E)})
\end{align}
To get good duality, we first assume that the $R$ charges of fundamental fields satisfy $R_Q+R_q=1-\frac{\Delta}{2}$. The above equation becomes
\begin{align}
\sum_j( t^{2R_Q+R_j^M}-t^{2-(2R_Q+R_j^M)})+\sum_l(t^{2R_Q+\Delta-R_l^E}-t^{2-(2R_Q+\Delta-R_l^E)})\,.
\end{align}
Now define $R_l^{'}=\Delta-R_l^E$, and the above equation would take the same form as (\ref{centralequation}). 
 Now the electric and magnetic gauge singlets are found from the set $\{R_j\}$ in the factorization as follows:
\begin{enumerate}
\item First find a subset $\{R_j^M\}$ from the solution, and the $R$ charge of the magnetic gauge singlet is given by $2R_Q+R_j^M$.

\item For the remaining elements in $\{R^E_j\}=\{R_j\}\backslash\{R_j^M\}$, the $R$ charge of the electric gauge singlet is given by $2R_q+\Delta-R_j^E$.  
\end{enumerate}
To make the duality work, one needs to add the superpotential term as follows: 
\begin{equation*}
W_E=\sum [M_I^E]M_{I^d}\,.
\end{equation*}
Here $M_{I^d}$ is the composite meson built from the electric fundamental degree of freedoms. This superpotential would couple the gauge singlet to the fundamental fields so that the electric theory is an irreducible theory.

\begin{figure}[h]
    \begin{center}

\tikzset{every picture/.style={line width=0.75pt}} 

\begin{tikzpicture}[x=0.55pt,y=0.55pt,yscale=-1,xscale=1]

\draw  [color={rgb, 255:red, 0; green, 0; blue, 0 }  ,draw opacity=1 ][fill={rgb, 255:red, 0; green, 0; blue, 0 }  ,fill opacity=1 ] (176.5,130) .. controls (176.5,128.21) and (175.04,126.75) .. (173.25,126.75) .. controls (171.46,126.75) and (170,128.21) .. (170,130) .. controls (170,131.79) and (171.46,133.25) .. (173.25,133.25) .. controls (175.04,133.25) and (176.5,131.79) .. (176.5,130) -- cycle ;
\draw  [color={rgb, 255:red, 0; green, 0; blue, 0 }  ,draw opacity=1 ][fill={rgb, 255:red, 0; green, 0; blue, 0 }  ,fill opacity=1 ] (273.25,126.75) .. controls (273.25,124.96) and (271.79,123.5) .. (270,123.5) .. controls (268.21,123.5) and (266.75,124.96) .. (266.75,126.75) .. controls (266.75,128.54) and (268.21,130) .. (270,130) .. controls (271.79,130) and (273.25,128.54) .. (273.25,126.75) -- cycle ;
\draw    (173.25,130) .. controls (194.5,107) and (242.5,100) .. (270,126.75) ;
\draw [shift={(226.36,109.54)}, rotate = 178.95] [fill={rgb, 255:red, 0; green, 0; blue, 0 }  ][line width=0.08]  [draw opacity=0] (8.93,-4.29) -- (0,0) -- (8.93,4.29) -- cycle    ;
\draw    (173.25,130) .. controls (175.5,145.75) and (245.5,156) .. (270,126.75) ;
\draw [shift={(215.16,145.31)}, rotate = 359.85] [fill={rgb, 255:red, 0; green, 0; blue, 0 }  ][line width=0.08]  [draw opacity=0] (8.93,-4.29) -- (0,0) -- (8.93,4.29) -- cycle    ;
\draw    (173.25,130) .. controls (117.75,113) and (166.5,53) .. (173.25,126.75) ;
\draw    (173.25,133.25) .. controls (166.5,188) and (115.5,173) .. (170,130) ;
\draw    (270,126.75) .. controls (245.5,67) and (289.5,62) .. (270,123.5) ;
\draw  [color={rgb, 255:red, 0; green, 0; blue, 0 }  ,draw opacity=1 ][fill={rgb, 255:red, 0; green, 0; blue, 0 }  ,fill opacity=1 ] (586.5,140) .. controls (586.5,138.21) and (585.04,136.75) .. (583.25,136.75) .. controls (581.46,136.75) and (580,138.21) .. (580,140) .. controls (580,141.79) and (581.46,143.25) .. (583.25,143.25) .. controls (585.04,143.25) and (586.5,141.79) .. (586.5,140) -- cycle ;
\draw  [color={rgb, 255:red, 0; green, 0; blue, 0 }  ,draw opacity=1 ][fill={rgb, 255:red, 0; green, 0; blue, 0 }  ,fill opacity=1 ] (683.25,136.75) .. controls (683.25,134.96) and (681.79,133.5) .. (680,133.5) .. controls (678.21,133.5) and (676.75,134.96) ..(676.75,136.75) .. controls (676.75,138.54) and (678.21,140) .. (680,140) .. controls (681.79,140) and (683.25,138.54) .. (683.25,136.75) -- cycle ;
\draw    (583.25,140) .. controls (604.5,117) and (652.5,110) .. (680,136.75) ;
\draw [shift={(624.96,120.14)}, rotate = 354.97] [fill={rgb, 255:red, 0; green, 0; blue, 0 }  ][line width=0.08]  [draw opacity=0] (8.93,-4.29) -- (0,0) -- (8.93,4.29) -- cycle    ;
\draw    (583.25,140) .. controls (585.5,155.75) and (655.5,166) .. (680,136.75) ;
\draw [shift={(636.65,155.02)}, rotate = 177.05] [fill={rgb, 255:red, 0; green, 0; blue, 0 }  ][line width=0.08]  [draw opacity=0] (8.93,-4.29) -- (0,0) -- (8.93,4.29) -- cycle    ;
\draw    (583.25,140) .. controls (527.75,123) and (576.5,63) .. (583.25,136.75) ;
\draw    (583.25,143.25) .. controls (541.5,189) and (525.5,183) .. (580,140) ;
\draw    (680,136.75) .. controls (655.5,77) and (699.5,72) .. (680,133.5) ;
\draw    (580,140) .. controls (502.5,151) and (516.75,127) .. (583.25,140) ;
\draw    (360,130) -- (438,130) ;
\draw [shift={(440,130)}, rotate = 180] [color={rgb, 255:red, 0; green, 0; blue, 0 }  ][line width=0.75]    (10.93,-3.29) .. controls (6.95,-1.4) and (3.31,-0.3) .. (0,0) .. controls (3.31,0.3) and (6.95,1.4) .. (10.93,3.29)   ;
\draw  [color={rgb, 255:red, 0; green, 0; blue, 0 }  ,draw opacity=1 ][fill={rgb, 255:red, 0; green, 0; blue, 0 }  ,fill opacity=1 ] (596.5,390) .. controls (596.5,388.21) and (595.04,386.75) .. (593.25,386.75) .. controls (591.46,386.75) and (590,388.21) .. (590,390) .. controls (590,391.79) and (591.46,393.25) .. (593.25,393.25) .. controls (595.04,393.25) and (596.5,391.79) .. (596.5,390) -- cycle ;
\draw  [color={rgb, 255:red, 0; green, 0; blue, 0 }  ,draw opacity=1 ][fill={rgb, 255:red, 0; green, 0; blue, 0 }  ,fill opacity=1 ] (693.25,386.75) .. controls (693.25,384.96) and (691.79,383.5) .. (690,383.5) .. controls (688.21,383.5) and (686.75,384.96) .. (686.75,386.75) .. controls (686.75,388.54) and (688.21,390) .. (690,390) .. controls (691.79,390) and (693.25,388.54) .. (693.25,386.75) -- cycle ;
\draw    (593.25,390) .. controls (614.5,367) and (662.5,360) .. (690,386.75) ;
\draw [shift={(634.96,370.14)}, rotate = 354.97] [fill={rgb, 255:red, 0; green, 0; blue, 0 }  ][line width=0.08]  [draw opacity=0] (8.93,-4.29) -- (0,0) -- (8.93,4.29) -- cycle    ;
\draw    (593.25,390) .. controls (595.5,405.75) and (665.5,416) .. (690,386.75) ;
\draw [shift={(646.65,405.02)}, rotate = 177.05] [fill={rgb, 255:red, 0; green, 0; blue, 0 }  ][line width=0.08]  [draw opacity=0] (8.93,-4.29) -- (0,0) -- (8.93,4.29) -- cycle    ;
\draw    (690,386.75) .. controls (665.5,327) and (709.5,322) .. (690,383.5) ;
\draw    (590,390) .. controls (512.5,401) and (526.75,377) .. (593.25,390) ;
\draw    (593.25,390) .. controls (484.5,433.75) and (520.5,395) .. (596.5,390) ;
\draw    (630,210) -- (630,308) ;
\draw [shift={(630,310)}, rotate = 270] [color={rgb, 255:red, 0; green, 0; blue, 0 }  ][line width=0.75]    (10.93,-3.29) .. controls (6.95,-1.4) and (3.31,-0.3) .. (0,0) .. controls (3.31,0.3) and (6.95,1.4) .. (10.93,3.29)   ;

\draw (111,65.4) node [anchor=north west][inner sep=0.75pt]    {$\left[ c^{*} u^{2} d^{*}\right]$};
\draw (101,175.4) node [anchor=north west][inner sep=0.75pt]    {$\left[ c^{*} u^{3} d^*\right]$};
\draw (227,92.4) node [anchor=north west][inner sep=0.75pt]    {$c$};
\draw (221,152.4) node [anchor=north west][inner sep=0.75pt]    {$d$};
\draw (267,52.4) node [anchor=north west][inner sep=0.75pt]    {$u$};
\draw (521,75.4) node [anchor=north west][inner sep=0.75pt]    {$\left[ c^{*} u^{2} d^{*}\right]$};
\draw (534,202.4) node [anchor=north west][inner sep=0.75pt]    {$\left[ cu^{3} d\right]$};
\draw (623,92.4) node [anchor=north west][inner sep=0.75pt]    {$c^{*}$};
\draw (622,162.4) node [anchor=north west][inner sep=0.75pt]    {$d^{*}$};
\draw (677,62.4) node [anchor=north west][inner sep=0.75pt]    {$u$};
\draw (482,152.4) node [anchor=north west][inner sep=0.75pt]    {$[ cd]$};
\draw (478,200.4) node [anchor=north west][inner sep=0.75pt]    {$\left[ cu^{2} d\right]$};
\draw (633,342.4) node [anchor=north west][inner sep=0.75pt]    {$c^{*}$};
\draw (632,412.4) node [anchor=north west][inner sep=0.75pt]    {$d^{*}$};
\draw (687,312.4) node [anchor=north west][inner sep=0.75pt]    {$u$};
\draw (473,364.4) node [anchor=north west][inner sep=0.75pt]    {$\left[ cu^{2} d\right]$};
\draw (467,409.4) node [anchor=north west][inner sep=0.75pt]    {$\left[ cu^{3} d\right]$};
\draw (378,102.4) node [anchor=north west][inner sep=0.75pt]    {$duality$};
\draw (647,242.4) node [anchor=north west][inner sep=0.75pt]    {$Reduction$};
\draw (461,122.4) node [anchor=north west][inner sep=0.75pt]    {$\left[ c^{*} u^{3} d^{*}\right]$};
\draw (476,175.4) node [anchor=north west][inner sep=0.75pt]    {$[ cud]$};

\end{tikzpicture}
\end{center}
\caption{Duality for superpotential $f=u^5$ with magnetic part of gauge singlets $R_j^M=\{u^2,u^3\}$ and electric part $R_j^E=\{1,u\}$.}
\label{more}
\end{figure}
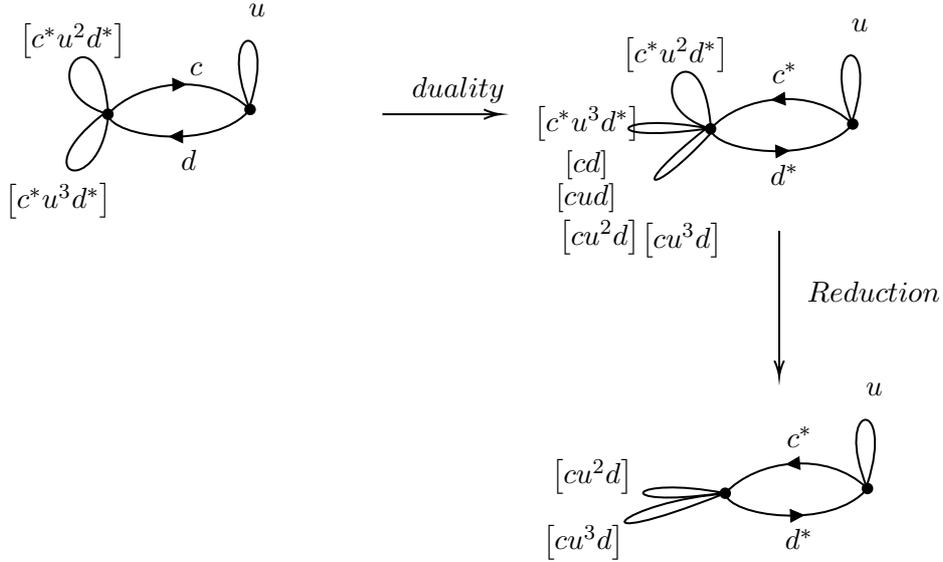

\textbf{Example}: Let's take $f=u^5$. The set of basic mesonic $R$ charges is $R_j=\{1,u,u^2, u^3\}$. Let's take the magnetic part of gauge singlets as $R_j^M=\{u^2,u^3\}$, and electric part as $R_j^E=\{1,u\}$. So the electric gauge singlets are $[c^*u^3d^*]$ and $[c^*u^2d^*]$ (We use these letters to indicate their $R$ charges). An electric superpotential is 
\begin{equation*}
W_E=[c^*u^3d^*]cd+[c^*u^2d^*]cud,
\end{equation*}
so that the singlet is coupled with the gauge theory. Now let's do duality following the general rule, and there would be four more new magnetic mesons in the magnetic 
frame. The superpotential takes the form
\begin{equation*}
W_M=[c^*u^3d^*][cd]+[c^*u^2d^*][cud]+\ldots\,.
\end{equation*}
The singlets $[c^*u^3d^*],[cd],[c^*u^2d^*],[cud]$ become massive. So we are left with two magnetic gauge singlets $[cu^2d],[cu^3d]$, which agrees with our setup. See Figure \ref{more} for the illustration of this duality model.

\textbf{Self-dual model}: It is now possible to have self-dual model, i.e. the dual gauge group and matter contents are the same as the original theory. The condition on $N_f, N_c$ is $aN_f-N_c=N_c$. We also have the same $R$
charge for the bi-fundamental fields: $R(c)=R(c^*)=\frac{1}{2}(1-\frac{\Delta}{2})$. We also arrange the electric
and magnetic gauge singlets symmetrically. Then we have the following possibilities: a) $A_1: \Delta=0$, and so $R(c)=R(c^*)=\frac{1}{2}$, $N_f=2N_c$; b) $A_k: \Delta=\frac{2k-2}{k+1}$, and so $R(c)=R(c^*)=\frac{1}{k+1}$, and $N_f={2N_c\over k}$; The gauge singlets are $[cd],\ldots, [cu^{[\frac{k-1}{2}]} d]$. The superpotential may be used to project some mesons out. 

\subsection{General solutions}
\label{generalcentral}
Every factorization of $y^n-1$ would give rise to a possible duality pair: one has complete electric and magnetic descriptions.  However, one often could not 
find a superpotential so that the truncation is derived from the equation of motion of it. One should also be aware that even for $E_7$  and $D_k$ ($k$ even) model, one needs to 
truncate the meson spectrum by hand \cite{Brodie:1996vx,Kutasov:2014yqa}, namely a quantum constraint is needed. However, there is little if any understanding about how such quantum constraint arises.

The author of \cite{Bajc:2019vbp} proposed superpotential for some models from the factorization. However, the potential does just reflect
the $R$ charges of adjoint fields, and  the proposed superpotential cannot give the truncation by 
using the equation of motion. Probably one should not take such superpotential too seriously.

Our point of view is that one should not be constrained by 
the existence of superpotential. Instead, 
the gauge theory studied above might be just part of a larger 
gauge theory, and the adjoint fields $u$s could be the composite of other fundamental fields.
The truncation on the mesonic spectrum could be due to some unknown quantum constraints. Of course,
it would be interesting to identify specific models with these more general solutions. In any case, we just take the specified mesonic spectrum as the input of our theory.

\begin{figure}[H]
	\centering
       \begin{tikzcd}
       |[draw,rectangle]| N_f\ar[r,bend left=10,"c"]&    
       |[draw,circle]| N_c \ar[l,bend left=10,"d"]
       \end{tikzcd}
       \begin{tikzcd}
		\stackrel{}{\Longrightarrow}    
	\end{tikzcd}       
    \begin{tikzcd}
       |[draw,rectangle]| N_f\ar[r,bend right=10,"d^*"']\ar[loop,out =200,in=230,looseness=9,"{[cd]}"',no head]\ar[loop,out =110,in=70,looseness=9,"{[cu^2d]}",no head]& 
       |[draw,circle]|  N_c^{'} \ar[l,bend right=10,"c^*"']
       \end{tikzcd}
       \caption{A duality for a special set of the mesonic spectrum derived from factorization $y^4-1=(y^2-1)(y^2+1)$.}
        \label{special}
\end{figure}
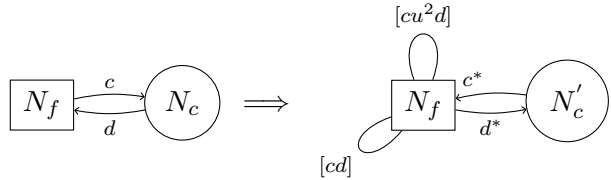

\textbf{Example}: Let's discuss a model derived from one factorization, and we will show how the duality in this situation would work. Let's
take the factorization $y^4-1=(y^2-1)(y^2+1)$, and so $\Delta=2$
 and there is no adjoint field, and $R_j=\{0,2\}$. It is not straightforward to form the mesons from the elementary field. However, we may add an adjoint field $u$ with $R$ charge $1$, and now the set of mesons is assumed to be $cd,cu^2d$. The dual theory works similarly, and a dual superpotential is formed:
 \begin{equation*}
W=[cd]c^*u^2d^*+[cu^2d]c^*d^*\,.
 \end{equation*}
 In the dual theory, the composite meson $c^*u^2 d^*,~c^*d^*$  are not in the chiral ring. 
 We then have the duality as shown in Figure \ref{special}. The rank of the dual gauge group is $2N_f-N_c$.

We have verified that the chiral spectrum of the dual pair matches, see the discussion at the end of section \ref{general}. Let's now verify that for the general duality proposal based on the factorization of the polynomial $y^n-1$. 
The  $\Tr R$ and $\Tr R^3$ anomalies match, which would ensure that the central charges match \cite{Anselmi:1997am}. The anomaly vanishing conditions of the original and dual gauge theories are (here we take the gauge group as $U(N_c)$ without losing any generality): 
\begin{equation*}
\begin{aligned}
  &N_c[\sum_u (R_u-1)+1]+(R_c-1)N_f=0,\\
  &N_c^{\prime}[\sum_u (R_u-1)+1]+(R_{c^*}-1)N_f=0\,.\\ 
  \end{aligned}
\end{equation*}

The original and dual R anomalies are
\begin{equation}
\begin{aligned}
\Tr R&=N_c^2[\sum_u (R_u-1)+1]+2N_cN_f(R_c-1)=N_cN_f(R_c-1),\\ 
\Tr R_{\text{dual}}&=N_c^{\prime2}[\sum_u (R_u-1)+1]+2 N_c^{\prime}N_f(R_{c^*}-1)+N_f^2\sum_j [(2R_c+R_j)-1]\\
&=N_c^{'}N_f (R_{c^*}-1)+N_f^2[2aR_c+{\Delta\over 2}a-a]\,.\\
\end{aligned}
\label{trR}
\end{equation}
Substitute $N_c^{'}=aN_f-N_c,~R_{c^*}=1-R_c-{\Delta\over 2}$, we have
\begin{align*}
\Tr R_{\text{dual}}&=(aN_f-N_c)N_f[-R_c-{\Delta\over 2}]+N_f^2[2aR_c+{\Delta\over 2}a-a] \nonumber\\
&=N_cN_f(R_c-1)+a N_f^2 (R_c-1) +\frac{\Delta N_c N_f}{2}+N_cN_f \nonumber\\
&=N_cN_f(R_c-1)+N_cN_f[1+\frac{\Delta}{2}-a[\sum_u(R_u-1)+1]] \nonumber\\
&=N_cN_f(R_c-1)\,.
\end{align*}
Here we used the equation $R_c=1-[\sum_u(R_u-1)+1]{N_c\over N_f}$, and the equation of $a$ (\ref{aequation}).

The original and dual $R^3$ anomalies are
\begin{equation*}
\begin{aligned}
\Tr R^3&=N_c^2[\sum_u (R_u-1)^3+1]+2N_cN_f(R_c-1)^3 \\
&=N_c^2[\sum_u (R_u-1)^3+1]-{2N_c^4\over N_f^2}(\sum_u(R_u-1)+1)^3,\\
\Tr R_{\text{dual}}^3&=N_c^{\prime2}[\sum_u (R_u-1)^3+1]+2 N_c^{\prime}N_f(R_{c^*}-1)^3+N_f^2\sum_j [(2R_c+R_j-1)^3]\,.
\end{aligned}
\end{equation*}
Substitute $N_c^{'}=aN_f-N_c,~R_{c^*}=1-R_c-{\Delta\over 2}$.
Define $x=\sum_u(R_u-1)^3+1,~y=\sum_j(R_j-1),~z=\sum_j(R_j-1)^2,~w=\sum_j(R_j-1)^3$. If these numbers satisfy the following equation:
\begin{equation}
\boxed{y=\frac{1}{2} a (\Delta-2),~~z=\frac{a \left(-2 a x+\Delta^3+8\right)}{3 (\Delta+2)},~~w=\frac{a \left(-4 a (\Delta-2) x+\Delta^4-16\right)}{4 (\Delta+2)}},
\end{equation}
then we also have the equality of the cubic anomalies. We have verified the above equality for many factorizations of 
polynomial $y^n-1$ and it would be interesting to prove the above identity exactly.

Let's check that other global anomalies also match. It is easy to check that $\Tr SU(N_f)^3=0$ and $\Tr SU(N_F) R^2=0$ agree for both sides. The last global anomaly that remains to check is 
\begin{equation*}
\Tr SU(N_f)^2R=\sum_i (R_i-1)\Tr SU(N_f)^2=\sum_i (R_i-1) ind.
\end{equation*}
Here $ind$ is the Dynkin index for the representation under the flavor symmetry. The original and dual $\Tr SU(N_f)^2R $ anomalies are
\begin{equation*}
    \begin{aligned}
&\Tr RFF=2N_c(R_c-1)\frac{1}{2}=N_c(R_c-1);\\
&\Tr R_{\text{dual}}F_{\text{dual}}F_{\text{dual}}=2N_c^{\prime}(R_{c^*}-1)\frac{1}{2}+N_f\sum_j(2R_c+R_j-1)\\
&=N_c^{\prime}(R_{c^*}-1)+a(\Delta/2+2R_c-1)N_f \,.
    \end{aligned}
\end{equation*}
One can check that they are equal from the equality of $\Tr R$ anomaly, see (\ref{trR}).

There is also a baryon number $U(1)_B$ symmetry.  The baryonic charge for quarks $c_i$ and antiquarks $d_i$  is normalized to be $\pm 1$. Since the duality maps baryons $B^{l_1\cdots l_a}$ to $B^{\tilde{l_1}\cdots\tilde{l_a}}$ and $\sum_i l_i=N_c,~\sum_i \tilde{l}_i=N_c^{\prime}$. Therefore the dual quarks $c_i^*$ should have the normalized $U(1)_B$ charges $ \frac{N_c}{N_c^{\prime}}=\frac{N_c}{a N_f-N_c}$. 

It is easy to see $\Tr U(1)_BRR=\Tr U(1)_B^3=0$ in both electric and magnetic theories. The nonzero anomalies are $\Tr RU(1)_B^2$, and the computations are
\begin{equation*}
\begin{aligned}
&\Tr RU(1)_B^2=2N_f N_c (R_c-1)=-2N_c^2x,\\
&\Tr R_{\text{dual}}U(1)^2_{B}= 2N_f N_c^{\prime}(R_{c^*}-1)(\frac{N_c}{aN_f-N_c})^2\\
&=2N_f (aN_f-N_c)(-R_c-\frac{\Delta}{2})(\frac{N_c}{aN_f-N_c})^2\\
&=-2N_c^2x\,.
\end{aligned}
\end{equation*}
Here we used the formula $a= {d+2\over 2x},R_c=1-x  {N_c\over N_f}$, with $x=\sum_u (R_u-1)+1$.

\newpage
\section{Duality for simple gauge group}\label{simplegaugegp}
The analysis of duality for other simple gauge groups can be carried out in a similar way. The basic idea is the same as discussed in last subsection: the composite mesons of one theory are mapped to the gauge singlets of the dual theory, and often an extra dual superpotential is needed. 

To make the duality work, we also find that the $R$ charges for tensorial fields and the meson spectrum should also satisfy the equation (\ref{pair}). However, we often have more 
constraints, namely, sometimes a $\mathbb{Z}_2$ action is needed to further separate the mesons.

\subsection{Classical gauge group with adjoint chiral}

\begin{figure}[H]
	\centering
       \begin{tikzcd}
       |[draw,rectangle]| 2N_f\ar[r,"c",no head]&    
       |[draw,circle]| G \ar[loop,out =290,in=250,looseness=5,"v",no head]\ar[loop,out =110,in=70,looseness=5,"u",no head]
       \end{tikzcd}
       \begin{tikzcd}
		\stackrel{}{\Longrightarrow}    
	\end{tikzcd}       
    \begin{tikzcd}
       |[draw,rectangle]| 2N_f\ar[loop,out =290,in=250,looseness=9,"",no head,blue]\ar[loop,out =140,in=170,looseness=9,"",no head]\ar[loop,out =200,in=230,looseness=9,"",no head,red]\ar[loop,out =110,in=70,looseness=9,"{[M_I]}",no head]& 
       |[draw,circle]|  G^{'} \ar[l,"c^*"', no head]\ar[loop,out =290,in=250,looseness=5,"v^*",no head]\ar[loop,out =110,in=70,looseness=5,"u^*",no head]
       \end{tikzcd}
       \caption{Duality for classical gauge group with adjoint chiral. The gauge singlets in the dual theory could be in either symmetric or anti-symmetric representation of the flavor group.}
        \label{classicalwithadjchiral}
\end{figure}
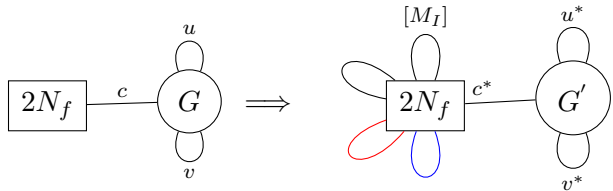

\begin{table}[H]
    \centering
    \begin{tabular}{|c|c|} \hline
        Gauge group & Dual gauge group \\ \hline
         $SU(N_c)$& $SU(aN_f-N_c)$ \\ \hline 
                  $SO(2N_c)$& $SO(2aN_f-2N_c+4)$ \\ \hline 
         $SO(2N_c+1)$& $SO(2aN_f-2N_c+3)$ \\ \hline 
         $USp(2N_c)$& $USp(2aN_f-2N_c-4)$ \\ \hline 
    \end{tabular}
    \caption{Gauge groups of the original theory and dual theory.}
    \label{classical}
\end{table}

Let's now assume that the gauge group is of the classical type, and the matter is in fundamental 
or adjoint representation. The main difference is that the matter transforms in the real (or pseudo-real) representation of the gauge group, 
so one does not need to add an arrow for the matter in fundamental representation. The original theory 
and its dual are shown in Figure \ref{classicalwithadjchiral}. The dual gauge group is determined by using anomaly free condition for $U(1)_R$ charge:
\begin{align*}
&h^\vee([u]-1+[v]-1)+N_f^*(R(c)-1)+h^\vee=0 , \nonumber\\
&h^{'\vee}([u]-1+[v]-1)+N_f^*(R(c^*)-1)+h^{'\vee}=0\,.  \nonumber\\
\end{align*}
Here $h^\vee$ is the dual Coxeter number.
Now assume that the adjoint chirals have a truncation, and so that the $R$ charges have a pairing $R_j+R_{j^{'}}=\Delta$. The existence of dual superpotential would relate the $R$ charges of the fundamental fields as $R(c)+R(c^*)=1-\frac{\Delta}{2}$. Here $N_f$ denotes the number of hypermultiplets (a pair of fundamental chiral fields).

The dual Coxeter number of the dual gauge group is then
\begin{equation*}
h^{\vee '}= aN_f^{*}-h^{\vee},
\end{equation*}
and $a$ is determined by the paring constant of mesons and $R$ charges of $[u], [v]$:
\begin{equation}
\boxed{a=\frac{2+\Delta}{2([u]+[v]-1)}\,.}
\end{equation}
And $N_f^* =N_f\times b$, here $N_f$ denotes the number of hypermultiplets (a pair of fundamental fields), and
$b=1$ for $A, C$ gauge groups, $b=2$ for $B,D$ gauge groups. The change of the gauge group is summarized in the Table \ref{classical}. 

To discuss the detailed duality, one needs to study the transformation behavior of the dressed meson under the flavor 
symmetry. 

\textbf{Example}: Let's consider $SO(N_c)$ type gauge group, and the fundamental representation is denoted as $V$. The adjoint representation is formed from anti-symmetric tensor product of V: $\wedge^2 V$. The basic invariant is $v^Tv$ built from fundamental representation $v$. The dressed mesons are formed as $v^TU_Iv$. When there is just one adjoint chiral $u$, it is easy to find the transformation behavior of the dressed meson under the flavor group: i.e. $v^Tu^iv$ is a symmetric (anti-symmetric) tensor of the flavor group if $i$ is even (odd). For the $Sp(N_c)$ gauge group, the adjoint representation is the symmetric power of the fundamental representation $Sym^2 V$, and
the  invariant is $v^TJ u^i v$, with  $J=\left[\begin{array}{cc}0&I_n\\-I_n& 0 \end{array} \right]$. It is then straightforward that $v^TJ u^i v$ is a symmetric (anti-symmetric) tensor of the flavor group if $i$ is even (odd). 

 For more than one adjoint chirals, to make duality work, one needs to define an involution on the set of mesons from the fundamental solutions: the meson is classified as positive type or negative type. There are $\frac{a+1}{2}$ ($\frac{a-1}{2}$)
 mesons with positive (negative) sign for $SO(N_c)$ gauge group. For $Sp(N_c)$ gauge group, there are $\frac{a-1}{2}$ ($\frac{a+1}{2}$)
 mesons with positive (negative) sign. Here $a$ is the total number of mesons.

We can also distribute electric and magnetic gauge singlets to get more dualities as we did for the $SU$ type theory in the last section.

The detailed computations for the $R$ anomalies can be found in the appendix \ref{adjointcentral}.

\subsection{General representations}
\begin{table}[H]
    \centering
    \begin{tabular}{|c|c|c|c|c|} \hline
        Group/ Index & Adjoint $=h^\vee$  & Symmetric  & Antisymmetric & Fundamental \\ \hline
         $SU(N)$& $N$  & $\frac{N}{2}+1$ & $\frac{N}{2}-1$ & $\frac{1}{2}$ \\  \hline
         $USp(2N)$& $N+1$ & Adj  & {$N-1$} & $\frac{1}{2}$ \\ \hline
       $SO(2N)$& $2N-2$ &{ $2N+2$ } & Adj & $1$ \\ \hline
      $SO(2N+1)$& $2N-1$ & {$2N+3$}  & Adj & $1$ \\ \hline
    \end{tabular}
    \caption{Dynkin indices for some common representations of the classical group. The representation besides the adjoint representation is not irreducible (often one needs to subtract a trivial representation to get an irreducible representation).}
    \label{indexrep}
\end{table}

Let's now consider a simple  gauge group coupled with general representations.
The representations are constrained so that there is no gauge anomaly (local gauge anomaly exists only for $SU(N)$
 gauge group \cite{banks1976comment}, and global anomaly exists for $Sp(2N)$ gauge group \cite{witten19822}). 

The anomaly free condition for $U(1)_R$ symmetry is 
\begin{align}
&\sum_{a}\mathrm{ind}(r_a)([u_a]-1)+\sum_i \mathrm{ind}(r_i)(R(r_i)-1)+h^\vee=0 \,. \nonumber
\end{align}
Here $r_a$ could be the general tensor (most often the two tensor) representation of gauge groups. The second sum is over the representation whose index is small, i.e. the defining representation. 
And $h^\vee$ is the dual Coxeter number. The index for some familiar representations of classical group is shown in Table \ref{indexrep}.

The duality often works as follows: First,  one assumes that the dual theory involves the conjugate representation of the electric theory,
and the gauge group is also the same type of the electric theory.
The similar anomaly free condition of the $U(1)_R$ charge of the dual theory is 
\begin{align}
&\sum_{a}\mathrm{ind}(r_a^*)([u_a]-1)+\sum_i \mathrm{ind}(r_i^*)(R(r_i^*)-1)+h^{'\vee}=0\,.  \nonumber
\end{align}
Secondly, the dual theory should have gauge singlets which are identified with the composite 
gauge invariant chiral mesons of the original theory. Therefore, one needs to study the invariant theory involving 
the fundamental fields and tensorial fields, which may be solved using the Weyl's invariant theory for the classical group case\cite{weyl1946classical}. The case of exceptional group would be discussed in next subsection.
The meson spectrum is formed by two vector fields, and the fields involving  tensor fields $u_a$. 
They would transform differently under the flavor group depending on the property of the tensor fields.

Once we find out the space of spectrum (again assuming the truncation of the meson spectrum to a finite set), 
a dual superpotential involving the gauge singlets is added so that the composite gauge invariant operators are projected out. The marginal condition for the added superpotential is then 
\begin{equation*}
2R(r_i)+2R(r_i^*)+R(U_I)+R(U_{I^d})=2\,.
\end{equation*}
Now assuming the basic meson involving tensorial field has the pairing $R(U_I)+R(U_{I^d})=\Delta$, and so 
\begin{equation*}
R(r_i^*)+R(r_i)=1-{\Delta\over2}\,.
\end{equation*}
We have the equation for the dual Coxeter numbers
\begin{align}
\boxed{h^\vee+\sum_a \mathrm{ind}(r_a)([u_a]-1)+h^{'\vee}+\sum_a \mathrm{ind}(r_a^*)([u_a]-1)+\sum_i \mathrm{ind}(r_i)(-1-\frac{\Delta}{2})=0\,.}
\label{dualcoxeter}
\end{align}
Here we assume that the index of the vector representation does not depend on the gauge group (this is 
true for the classical group, as we assume the dual gauge group is of the same type).

From this formula, it is clear that the dual gauge group is  determined by the $R$ charges of tensorial fields $[u_a]$, and the pairing constant $\Delta$. Here more data is needed, one needs to have an extra
involution on the set of mesons to indicate the transformation property of the dressed meson under the flavor symmetry group.

\textbf{Example}: Let's consider $SO(2N_c)$ coupled with a traceless symmetric tensor $X$ and $N_f$ hypermultiplets in fundamental representation, and a superpotential $\Tr X^{k+1}$ is added. The dual gauge group is found from above equations:
\begin{align*}
&2N_c-2+(2N_c+2)[\frac{2}{k+1}-1]+2N_c^{'}-2+(2N_c^{'}+2)[\frac{2}{k+1}-1]+2N_f(-1-\frac{k-1}{k+1})=0 , \nonumber\\
\Rightarrow~&2N_c^{'}=2kN_f+4k-2N_c\,.
\end{align*}
This duality has been studied in \cite{Intriligator:1995ax}. More general classes are discussed in appendix \ref{other}.

\textbf{General case}: In the above analysis, one assumes that dual gauge group is the same type as the original gauge group. It might be possible that the dual theory takes a different form. The duality works similarly: the composite gauge invariant operator of the electric theory is mapped to the elementary (gauge singlet) field of the dual theory. Often a superpotential is needed to project the composite chiral operator of the dual theory out. Again, a detailed knowledge of the chiral spectrum (such as the truncation to a finite set) is crucial to identify the dual theory. See \cite{Pouliot:1995zc} for some examples, and certainly more examples can be found by considering the truncation of tensorial fields discussed in last section.

\textbf{Self-dual example}: One may engineer the matter representations so that the gauge groups are not changed,
see \cite{Csaki:1997cu,Hook:2014ywa} for some examples. It seems possible to find more examples using our general analysis in last section.

\subsection{Exceptional gauge group}
For exceptional group, one mostly needs
to only consider the fundamental representation $F$ as other representations have large index. 
The basic idea of duality is still mapping the composite chiral operators in electric theory to 
the elementary field in the magnetic theory. The first question is to classify the chiral spectrum, which is the study of the invariant theory for exceptional group. As in the classical group case, one can have invariant built from a fundamental field and an anti-fundamental field (mesons), or the 
invariant built from the determinant (baryons). There are more possibilities for the exceptional groups, see \cite{Giddings:1995ns}. For example, one could have operators built from  three fundamental fields for $E_6$ group. This makes the study of duality more complicated, i.e. the dual gauge group is often not of the same type. 

We may consider the theory where the dual gauge theory has the same type as the electric theory. 
The duality should work in similar way: one adds gauge singlet $[M]$ for the composite meson in the electric theory, and 
adds a superpotential to project out the composite meson in the dual gauge theory. The dual gauge group might be found from the anomaly free condition of $U(1)_R$ charge, and the self-dual condition constraints the possible matter content of the electric theory.

\textbf{Example}: For $E_6$ group, the index of fundamental representation is $3$, and the index of adjoint representation is $12$. Let's assume that there are $n_f$ fundamentals $F$, and it is known that there is a gauge invariant term $F^3$. The duality changes the representation $F$ to the dual representation $F^*$. To find the dual gauge group, one  needs to modify the relation between the $R$ charges of the fundamental fields, as now the basic invariant is a cubic invariant. According to the duality proposal, there should be a superpotential term in 
the magnetic description
\begin{equation*}
W=[F^3]F^{*3}\,.
\end{equation*}
Here $[F^3]$ is the gauge singlet, therefore the relation between $R$ charges of $F$ and $F^*$ is changed as
\begin{equation*}
R(F)+R(F^*)=\frac{2}{3}\,.
\end{equation*}
This is in contrast with the classical group case (Compare (\ref{fundarcharge})). The Coxeter number of the dual gauge group takes the form
\begin{equation*}
h^{'\vee}=4 n_f-h^\vee\,.
\end{equation*}
When $n_f=6$, one has the self-dual situation $h^{\vee'}=h^\vee$: one can add a marginal sextic interaction in the electric theory, and the theory is completely symmetric for electric and magnetic description. In this model, although the $R$ charge of the fundamental fields is $\frac{1}{3}$ which is below the unitarity bound, the gauge invariant operator involves three fundamental fields and so its $R$ charge is $1$, which is above the unitarity bound. 
This model has been studied in \cite{Ramond:1996ku}. More self-dual examples based on exceptional gauge group were also studied in \cite{Distler:1996ub}.

Let's now add one more adjoint chiral field $u$ to above model, so that the set of dressed mesons is $\{Fu^iF^2\}$. The dual theory should have a superpotential 
\begin{equation*}
W=\sum[Fu^iF^2]F^{*}u^jF^{*2}\,.
\end{equation*}
To make the duality work, the undressed mesons should be paired as $R(u^i)+R(u^j)=\Delta$, and so $R_F+R_{F^{*}}=\frac{2}{3}-\frac{\Delta}{3}$. The dual Coxeter number of the dual gauge group is 
\begin{equation*}
h^{'\vee}=n_f{\Delta+4\over R(u)}-h^\vee=(3k+1)n_f-h^\vee\,.
\end{equation*}
Here we assume there is a $A_k$ type potential for adjoint field $u$.
It would be interesting to identify the dual theory for this more general model (notice that $k=1,4,5,7,8,11$).

Alternatively, We may   have $N_f$ pairs of fundamentals $F$ and anti-fundamentals $\bar{F}$, and an adjoint $u$ with $A_{k}$ type superpotential.  So there could be gauge invariant $Fu^k\bar{F}$ besides the cubic invariant $F^3, \bar{F}^3$. We now add similar dual potential as for the classical group case, and 
the data for the dual gauge group would be 
\begin{equation*}
h^{'\vee}=6k N_f-h^\vee\,.
\end{equation*}
When $k=1$ (no adjoint), one finds that $N_f=4$ so that the dual gauge group has the same dual Coxeter number. This 
example is different from that in \cite{Ramond:1996ku} as the duality maps gauge invariant $F\bar{F}$ to the gauge singlet here (the self-dual example in \cite{Ramond:1996ku} maps a gauge invariant $F^3$ to a gauge singlet). When $k=4, N_f=1$, we also have a self-dual model.

 \begin{table}[H]
    \centering
    \begin{tabular}{|c|c|c|} \hline
        Group/ Index & Adjoint $=h^\vee$  & Fundamental(\textbf{dim})index \\ \hline
        $G_2$ & 4  & (\textbf{7}) 1 \\ \hline
       $F_4$ & 9 & (\textbf{26})3 \\ \hline
       $E_6$ & 12 & (\textbf{27})3  \\ \hline
       $E_7$ & 18 & (\textbf{56})6 \\ \hline
       $E_8$ & 30 & (\textbf{248})30 \\ \hline
    \end{tabular}
    \caption{Dynkin indices and dimensions of the adjoint representation and fundamental representation of the exceptional groups. The Dynkin index of the adjoint representation is equal to the dual Coxeter number by our convention.}
    \label{tab:my_label}
\end{table}

\section{Duality for semi-simple gauge groups}\label{semisimple}
\begin{figure}[H]
	\begin{center}
\begin{tikzpicture}[every state/.style={minimum width={2cm} ,thick,align=center}]

\begin{scope}
    \path 
    (-1,0) node (g1) {$G_1$}
    (0,0) node (g2) {$G_2$};
    \node[draw,fit=(g1) (g2)] (g12) {};
    \node[draw] (n1) at (-2.5,-1.5) {$N_1$};
    \node[draw] (n2) at (1.5,-1.5) {$N_2$};
    \draw[->] (g12.220)--(n1.30);
    \draw[->] (g12.320)--(n2.150);
    \draw[->] (n1.50)--(g12.210);
    \draw[->] (n2.130)--(g12.330);
    \draw[->] (g1.10)--(g2.170);
    \draw[<-] (g1.-10)--(g2.-170);
\end{scope}

\begin{scope}[xshift=6cm]
    \path 
    (-1,0) node (g1) {$G_1^*$}
    (0,0) node (g2) {$G_2^*$};
    \node[draw,fit=(g1) (g2)] (g12) {};
    \node[draw] (n1) at (-2.5,-1.5) {$N_1$};
    \node[draw] (n2) at (1.5,-1.5) {$N_2$};
    \draw[->] (g12.220)--(n1.30);
    \draw[->] (g12.320)--(n2.150);
    \draw[->] (n1.50)--(g12.210);
    \draw[->] (n2.130)--(g12.330);
    \draw[->] (g1.10)--(g2.170);
    \draw[<-] (g1.-10)--(g2.-170);
    \draw[->] (n1.10)--(n2.170);
    \draw[<-] (n1.-10)--(n2.-170);
    \path[ every loop/.style={looseness=8, in=290, out=250}]
            (n1) edge [loop] node[above] {} ();
    \path[ every loop/.style={looseness=8, in=230, out=200}]
            (n1) edge [loop] node[above] {} ();
    \path[ every loop/.style={looseness=8, in=290, out=250}]
            (n2) edge [loop] node[above] {} ();
    \path[ every loop/.style={looseness=8, in=340, out=310}]
            (n2) edge [loop] node[above] {} ();
\end{scope}

\draw[black,->] (2,-0.5) -- (3,-0.5);

\end{tikzpicture}
\end{center}
\caption{The original theory has two gauge groups $G_1, G_2$ and two flavor groups denoted by $N_1, N_2$. There are bifundamental matters between two gauge groups and fundamental charged over one of the gauge group. Dual theory has mesons as adjoint representations of two flavor groups, and the bi-fundamental matter. The spectrum of gauge singlet is determined by the meson spectrum of original theory.
}
        \label{twogaugegpincore}
\end{figure}
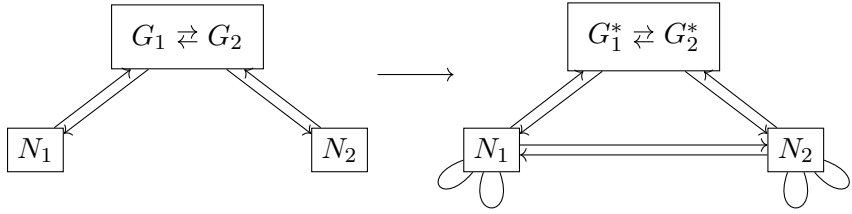

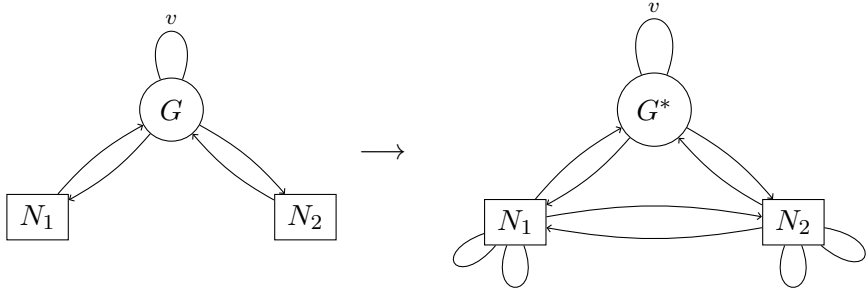
\begin{figure}[H]
	\centering
       \begin{tikzcd}
       &|[draw,circle]| G\ar[dl,bend left=10,""]\ar[dr,bend left=10,""]\ar[loop,out =110,in=70,looseness=8,"v",no head]&    \\
       |[draw,rectangle]| N_1\ar[ur,bend left=10,""] &
       & |[draw,rectangle]| N_2\ar[ul,bend left=10,""]
       \end{tikzcd}
       \begin{tikzcd}
		\stackrel{}{\longrightarrow}    
	\end{tikzcd}       
    \begin{tikzcd}
      &|[draw,circle]| G^*\ar[dl,bend left=10,""]\ar[dr,bend left=10,""]\ar[loop,out =110,in=70,looseness=8,"v",no head]&    \\
       |[draw,rectangle]| N_1\ar[ur,bend left=10,""] \ar[rr,bend left=10,""] \ar[loop,out =290,in=250,looseness=9,"",no head]\ar[loop,out =230,in=200,looseness=9,"",no head]&
       & |[draw,rectangle]| N_2\ar[ul,bend left=10,""]\ar[ll,bend left=10,""]\ar[loop,out =290,in=250,looseness=9,"",no head]\ar[loop,out =350,in=320,looseness=9,"",no head]
       \end{tikzcd}
       \caption{The original theory has one gauge group $G$ and two flavor groups denoted by $N_1, N_2$ which could be further gauged.  Dual theory has mesons in adjoint representations of $U(N_1)$ and $U(N_2)$ group, and there are also new fields in bifundamental representations of  $U(N_1)$ and $U(N_2)$ group.}
        \label{quivergaugethy}
\end{figure}

Let's now assume the gauge group is  semi-simple, i.e. the gauge group is a product of simple factors. 
There are two situations that one could consider: the first one regards several simple gauge groups as a single core 
, and there could be fields such as fundamental representations charged with individual gauge groups, and the duality acts on the gauge groups in the core simultaneously, see \cite{Intriligator:1995ax} for examples;
the second one is more common: one could consider quiver gauge theory where the duality acts on each simple gauge group separately, see \cite{Xie:2012mr, Fang:2023wxa} for examples.
The duality works in the similar way: the dual gauge groups are changed, and one adds the gauge singlet in the dual theory and a superpotential to project out the composite mesons. Again, the crucial condition is that the set of mesons is truncated, and there is a paring between the meson spectrum. 
See Figure \ref{twogaugegpincore} for the illustration of an example whose core has more than one gauge groups, 
and Figure \ref{quivergaugethy} for the quiver gauge theory.

Using the result of last sections, it is possible to have 
a lot more interesting dualities by considering other truncations. Some examples would be discussed in \cite{Fang:2024}.

\section{Conclusion}\label{conclusion}
We discussed the general picture of duality of $\mathcal{N}=1$
non-abelian gauge theory. The main point is that the spectrum of undressed mesons 
should take particular form to make the duality work. The spectrum is elegantly 
encoded in the factorization of polynomial $y^n-1$ into a positive and a negative part $y^n-1=\Phi_{+}\Phi_{-}$. 
It is interesting to find physical models with those spectrum.

We mainly discussed the combinatorial nature of duality, and it would be interesting 
to study the dynamical consequences of the dualities. Notice that the duality often works
even if there are subtle dynamical questions such as the appearance of accidental symmetry, 
and the dual description could be useful to answer some of the difficult dynamical questions.

In this paper, we only study the duality of gauge theory coupled with free matters. 
It would be interesting to generalize the duality to  $\mathcal{N}=1$ gauge theory coupled 
with strongly coupled matters. Many interesting $\mathcal{N}=1$ theories can be formed by 
relevant deformation of $\mathcal{N}=2$ theories \cite{Xie:2019aft}, and it is interesting to generalize the consideration of this paper to those more general cases. It is quite likely that the unusual 
meson spectrum studied in this paper could be realized in those models.

It would be interesting to generalize the consideration of this paper to 3d $\mathcal{N}=2$ theories.

\section*{Acknowledgments}
The work of Dan Xie is supported by national key research
and development program of China (NO. 2020YFA0713000).

\newpage
\appendix

\section{Gauge invariant operators}
We summarize the results of the invariant theory of groups, from which one can get the generators for 
the gauge invariant operators for a single gauge group. For more details, see \cite{popov2012algebraic}.
The basic invariants for the classical gauge groups with fundamental representations are: 
\begin{enumerate}
\item $SU(n)$  group: one 
needs to have both the fields $\phi$ in fundamental representation and $u$ in anti-fundamental representations; The invariants are $(u,\phi)$ and the bracket denotes the Hermitian form; and we also have the determinantal invariant $\det(u_1,u_2,\ldots, u_n)$, $\det(\phi_1,\phi_2,\ldots, \phi_n)$. 
\item $SO(n)$ group: $v_i^Tv_j$ and $\det(v_1,v_2,\ldots, v_n)$.
\item $Sp(n)$ group: $v_i^TJv_j$ with $J$ the 
matrix $J=\left[\begin{array}{cc}0&I_n\\-I_n& 0 \end{array} \right]$. 
\end{enumerate}
There are also relations between these invariants, and they are summarized in Table \ref{relation}.

\begin{table}[h]
    \centering
    \begin{tabular}{|c|l|} \hline
         $SU(n)$ & $\det((u_i,\phi_j))_{i,j=1}^n=\det(u_1,\ldots, u_n)\det(\phi_1,\ldots, \phi_n)$, \\ 
         ~&  $\sum_{i}(-1)^i\det(u_0,\ldots,\hat{u_i},\ldots,u_n)(u_i,\phi)=0$,     \\ 
                  ~&  $\sum_{i}(-1)^i\det(\phi_0,\ldots,\hat{\phi_i},\ldots,\phi_n)(u,\phi_i)=0$,     \\ 
   ~&  $\sum_{i}(-1)^i\det(u_0,\ldots,\hat{u_i},\ldots,u_n)\det(u_i,v_1,\ldots, v_{n-1})=0$,    \\ 
                     ~&  $\sum_{i}(-1)^i\det(\phi_0,\ldots,\hat{\phi_i},\ldots,\phi_n)\det(\phi_i,\psi_1,\ldots, \psi_{n-1})=0$.   \\ \hline
         $SO(n)$ &$\det((u_i,v_j))_{i,j=0}^n=0$,       \\ 
      & $\det((u_i,v_j))_{i,j=1}^n=\det(u_1,\ldots, u_n)\det(v_1,\ldots, v_n)$. \\ \hline 
               $Sp(n)$ &$Pf((u_i,u_j))_{i,j=1}^{n+2}=0$.      \\ \hline

    \end{tabular}
    \caption{The typical relations for the invariants of classical group. Here the hat over a vector means the omission from the sequence. The Plaffian of a matrix $A$ is defined as $Pf(A)^2=\det A$. }
    \label{relation}
\end{table}

When there are tensor fields, one can build the invariants by first contracting the index and then forming the basic invariant. One uses metric $g$ and $g^{-1}$ in the $SO$ type case to raise or lower the index, and anti-symmetric form $w$ ($w^{-1}$) in $Sp$ group case to raise or lower the index.

The invariant tensors for exceptional groups are listed in Table \ref{exceptional}, see also \cite{Giddings:1995ns}.
The relations for these invariants are more complicated.

\begin{table}[H]
    \centering
    \begin{tabular}{|c|c|} \hline
    Lie algebra & Invariant tensor \\ \hline
         $G_2$& $\delta^{ab},f^{abc}$ \\ \hline
             $F_4$& $\delta^{ab},d^{abc}$ \\ \hline
         $E_6$& $d^{abc}$ \\ \hline
         $E_7$& $f^{abc},d^{abc}$ \\ \hline
         $E_8$& $\delta^{AB},f^{ABC}$ \\ \hline
    \end{tabular}
    \caption{Invariant tensors for fundamental representation of the exceptional group. Here $d$ is the symmetric tensor, $f$ is the anti-symmetric tensor.}
    \label{exceptional}
\end{table}

\section{Anomaly computations for classical groups}
In this appendix, we will compute the $\Tr R$ and $\Tr R^3$ anomalies for gauge theory with 
classical gauge group.  

\subsection{Adjoint matter}
\label{adjointcentral}
Let's do anomaly computation for the duality of classical gauge groups. We assume that the tensorial fields 
are all  adjoint fields.

First, we have the anomaly free condition for 
$U(1)_R$ symmetry:
\begin{equation*}
\begin{aligned}
  &h^\vee[\sum_u (R_u-1)+1]+(R_c-1)N_f^*=0,\\
  &h^{'\vee}[\sum_u (R_u-1)+1]+(R_{c^*}-1)N_f^*=0\,.\\ 
  \end{aligned}
\end{equation*}

\textbf{B,D} gauge group: For $SO(N_c)$ gauge group, define $x=\sum_u(R_u-1)+1$. We have 
\begin{equation*}
\begin{aligned}
\Tr R&=x \dim G+2N_f N_c(R_c-1),\\ 
\Tr R_{\text{dual}}&=x \dim G^{'}+2N_f N_c^{'}(R_{c^*}-1)+(2N_f^2+N_f)\sum_{j~\text{even}} [(2R_c+R_j)-1]\\&+ (2 N_f^2-N_f)\sum_{j~\text{odd}} [(2R_c+R_j)-1]\,.\\
\end{aligned}
\end{equation*}
 Now assume that one can define a parity on the set of adjoint fields, and so the set of mesons built from adjoint also has a parity: there are ${a-1\over 2}$ odd mesons (anti-symmetric representation of $Sp$ flavor group), and $\frac{a+1}{2}$ even mesons (symmetric representation of $Sp$ flavor group). We have 
\begin{equation*}
\begin{aligned}
\Tr R_{\text{dual}}&=x \dim G^{'}+2N_f N_c^{'}(R_{c^*}-1)+(2N_f^2+N_f) [(2R_c \frac{a+1}{2}+\Delta\frac{a+1}{4})-\frac{a+1}{2}]\\&+ (2 N_f^2-N_f) [(2R_c \frac{a-1}{2}+\Delta\frac{a-1}{4})-\frac{a-1}{2}]\\
&=x \dim G^{'}+2N_f N_c^{'}(R_{c^*}-1)+2 N_f (2 a N_f+1)[R_c+\frac{\Delta}{4}-\frac{1}{2}]\,.
\end{aligned}
\end{equation*}
Here $\dim G= \frac{1}{2}(N_c^2-N_c)$, and 
$N_c^{'}=2a N_f-N_c+4,~~R_{c^*}-1=-R_c-\frac{\Delta}{2}$. 
Using $R_c= 1-\frac{(N_c-2) x}{2 N_f},a=\frac{\Delta+2}{2 x}$, one can finally check that $\Tr R_{\text{dual}}=\Tr R$.

Next we check that $\Tr R^3= \Tr R^3_{\text{dual}}$.
The original and dual $R^3$ anomalies are
\begin{equation*}
\begin{aligned}
\Tr R^3&=\frac{1}{2}(N_c^2-N_c)[\sum_u (R_u-1)^3+1]+2N_cN_f(R_c-1)^3, \\
 \Tr R^3_{\text{dual}}&=\frac{1}{2}(N_c^{\prime 2}-N_c^{\prime})[\sum_u (R_u-1)^3+1]+2 N_c^{\prime}N_f(R_{c^*}-1)^3\\
&+(2N_f^2+N_f)\sum_{j~\text{even}} [(2R_c+R_j)-1]^3+ (2 N_f^2-N_f)\sum_{j~\text{odd}} [(2R_c+R_j)-1]^3\,.\\
\end{aligned}
\end{equation*}
Substitute $N_c^{'}=2aN_f-N_c+4,~R_{c^*}=1-R_c-{\Delta\over 2}$.
Define $x=\sum_u(R_u-1)^3;~y_1=\sum_{j ~\text{odd}}(R_j-1),~y_2=\sum_{j ~\text{even}}(R_j-1);~z_1=\sum_{j~\text{odd}}(R_j-1)^2,~z_2=\sum_{j~\text{even}}(R_j-1)^2;~w_1=\sum_{j~\text{odd}}(R_j-1)^3,~w_2=\sum_{j~\text{even}}(R_j-1)^3$.
Then we have $\Tr R^3= \Tr R^3_{\text{dual}}$ if these numbers satisfy the following equations:
\begin{equation}\label{trr3}
\begin{aligned}    
&y_1= \frac{1}{4} (a-1) (\Delta-2),~y_2= \frac{1}{4} (a +1)(\Delta-2),\\
&z_1=-\frac{4 a^2 x+4 a^2-2 a \Delta^3-12 a x-28 a+3 \Delta^3+6 \Delta^2+12 \Delta+24}{12 (\Delta+2)},\\
&z_2= -\frac{4 a^2 x+4 a^2-2 a \Delta^3+12 a x-4 a-3 \Delta^3-6 \Delta^2-12 \Delta-24}{12 (\Delta+2)},\\
&w_1=-\frac{4 a^2 \Delta x+4 a^2 \Delta-8 a^2 x-8 a^2-a \Delta^4-12 a \Delta x-12 a \Delta+24 a x+40 a+2 \Delta^4+4 \Delta^3-16 \Delta-32}{8 (\Delta+2)}, \\
&w_2= -\frac{4 a^2 \Delta x+4 a^2 \Delta-8 a^2 x-8 a^2-a \Delta^4+12 a \Delta x+12 a \Delta-24 a x-8 a-2 \Delta^4-4 \Delta^3+16 \Delta+32}{8 (\Delta+2)}\,.
\end{aligned}
\end{equation}
\bigskip

One can check that other global anomalies also match.

\textbf{C} gauge group:
For $Sp(N_c)$ gauge group, define $x=\sum_u(R_u-1)+1$. We have 
\begin{equation*}
\begin{aligned}
\Tr R&=x \dim G+2N_f  N_c(R_c-1),\\ 
\Tr R_{\text{dual}}&=x \dim G^{'}+2N_f N_c^{'}(R_{c^*}-1)+(2N_f^2-N_f)\sum_{j~\text{even}} [(2R_c+R_j)-1]\\&+ (2 N_f^2+N_f)\sum_{j~\text{odd}} [(2R_c+R_j)-1]\,.\\
\end{aligned}
\end{equation*}
Now assume that one can define a parity on the set of adjoint fields, and so the set of mesons built from adjoint also has a parity: there are ${a-1\over 2}$ odd mesons (symmetric representation of flavor group), and $\frac{a+1}{2}$ even mesons (anti-symmetric representation of flavor group). We have 
\begin{equation*}
\begin{aligned}
\Tr R_{\text{dual}}&=x \dim G^{'}+2N_f N_c^{'}(R_{c^*}-1)+(2N_f^2-N_f) [(2R_c \frac{a+1}{2}+\Delta\frac{a+1}{4})-\frac{a+1}{2}]\\&+ (2 N_f^2+N_f) [(2R_c \frac{a-1}{2}+\Delta\frac{a-1}{4})-\frac{a-1}{2}]\\
&=x \dim G^{'}+2N_f N_c^{'}(R_{c^*}-1)+2 N_f (2 a N_f-1)[R_c+\frac{\Delta}{4}-\frac{1}{2}]\,.
\end{aligned}
\end{equation*}
We have  $\dim G= \frac{1}{2}(N_c^2+N_c)$, and 
$N_c^{'}=2a N_f-N_c-4,~R_{c^*}-1=-R_c-\frac{\Delta}{2}$. 
Using $R_c= 1-\frac{(N_c+2) x}{ 2N_f},~a=\frac{\Delta+2}{2 x}$, one can finally check that $\Tr R_{\text{dual}}=\Tr R$.

We can also check that $\Tr R^3= \Tr R^3_{\text{dual}}$ if these numbers $x,y_1,y_2,z_1,z_2,w_1,w_2$ satisfy the same equation \ref{trr3}.

\subsection{Other matter}
\label{other}

\subsubsection{\textbf{B,D} gauge group with symmetric matter} 
First, we have the anomaly free condition for 
$U(1)_R$ symmetry:
\begin{equation*}
\begin{aligned}
  &h^\vee+(h^\vee+4)[\sum_u (R_u-1)]+(R_c-1)N_f^*=0,\\
 & h^{'\vee}+(h^{\vee'}+4)[\sum_u (R_u-1)]+(R_{c^*}-1)N_f^*=0\,.\\ 
  \end{aligned}
\end{equation*}
So we have 
\begin{equation*}
h^{\vee'}=aN_f^*-h^\vee-\frac{8x-8}{x}\,.
\end{equation*}
Here $a={\Delta+2\over 2x}$, $x=\sum_u(R_u-1)+1$.
The anomalies are
\begin{equation*}
\begin{aligned}
\Tr R&=\frac{1}{2}(N_c^2-N_c)+\frac{1}{2}(N_c^2+N_c)\sum_u(R_u-1)+2N_f N_c(R_c-1),\\ 
\Tr R_{\text{dual}}&=\frac{1}{2}(N_c^{'2}-N_c^{'})+\frac{1}{2}(N_c^{'2}+N_c^{'})\sum_u(R_u-1)+2N_f N_c^{'}(R_{c^*}-1)\\&+(2N_f^2+N_f)\sum_{j \in I} [(2R_c+R_j)-1]+(2N_f^2-N_f)\sum_{j \notin I} [(2R_c+R_j)-1]\,.\\
\end{aligned}
\end{equation*}
For $A$ type potential, $I$ is the whole set of mesons. For $D$ type potential, $I$ denotes the combination $u^av^b$ such
that $ab$ is even.

More generally, assume that there are $m$ symmetric mesons, and $n$ antisymmetric mesons under the flavor symmetry. We may fix $m,n$ by matching the anomalies. 
First we need to have the match of $\Tr R$ anomaly: 
\begin{equation*}
\begin{aligned}
    \Tr R_{\text{dual}}&=\frac{1}{2}(N_c^{'2}-N_c^{'})+\frac{1}{2}(N_c^{'2}+N_c^{'})\sum_u(R_u-1)+2N_f N_c^{'}(R_{c^*}-1)\\&+(2N_f^2+N_f)(m)[2R_c+d/2-1]+(2N_f^2-N_f)(n) [(2R_c+d/2-1]\,.
\end{aligned}
\end{equation*}
The $\Tr RUSp(2N_f)^2$ anomalies in electric and magnetic frames are
\begin{equation*}
\begin{aligned}
 &\Tr RFF=\frac{1}{2} N_c (R_c-1),\\
 &\Tr R_{\text{dual}}FF=\left(\frac{d}{2}+2 R_c-1\right) (( m) (N_f+1)+( n) (N_f-1))+\frac{1}{2} N_c^{\prime} (R_{c^{*}}-1).
\end{aligned}
\end{equation*}
By solving $\Tr R=\Tr R_{\text{dual}}$ and $\Tr RFF=\Tr R_{\text{dual}}FF$ at the same time, we find $m,n$ : 
\begin{equation*}
\begin{aligned}
m= -\frac{-d+2 x-6}{4x}=\frac{a}{2}+\frac{4-2x}{4x},\\
n= -\frac{-d-2 x+2}{4 x}=\frac{a}{2}-\frac{4-2x}{4x}\,.
\end{aligned}
\end{equation*}

For $A_k$ type potential, $x=\frac{2}{k+1},~a=k$, we have $m=k=a,~n=0$.

For $D_k$ type potential, $x=\frac{1}{k+1},~a=3k$, we have $m=\frac{5k+1}{2},~n=\frac{k-1}{2}$. When $k$ is odd, it gives a possible duality. It is possible to find other solutions from the factorization of the polynomials $y^n-1$.

\subsubsection{\textbf{C} gauge group with anti-symmetric matter}
First, we have the anomaly free condition for 
$U(1)_R$ symmetry:
\begin{equation*}
\begin{aligned}
  &h^\vee+(h^\vee-2)[\sum_u (R_u-1)]+(R_c-1)N_f^*=0,\\
  &h^{'\vee}+(h^{\vee'}-2)[\sum_u (R_u-1)]+(R_{c^*}-1)N_f^*=0\,.\\ 
  \end{aligned}
\end{equation*}
So we have 
\begin{equation*}
\begin{aligned}
&h^{\vee'}=aN_f^*-h^\vee+\frac{4x-4}{x},\\
&R_c=1-\frac{(N_c+2)x-4(x-1)}{2N_f}\,.
\end{aligned}
\end{equation*}
Here $a={\Delta+2\over 2x}$, $x=\sum_u(R_u-1)+1$.
So $N_c^{'}=2aN_f-N_c+\frac{8x-8}{x}-4$.
The anomalies are
\begin{equation*}
\begin{aligned}
\Tr R&=\frac{1}{2}(N_c^2+N_c)+\frac{1}{2}(N_c^2-N_c)\sum_u(R_u-1)+2N_f N_c(R_c-1),\\ 
\Tr R_{\text{dual}}&=\frac{1}{2}(N_c^{'2}+N_c^{'})+\frac{1}{2}(N_c^{'2}-N_c^{'})\sum_u(R_u-1)+2N_f N_c^{'}(R_{c^*}-1)\\&+(2N_f^2-N_f)\sum_{j \in I} [(2R_c+R_j)-1]+(2N_f^2+N_f)\sum_{j \notin I} [(2R_c+R_j)-1]\,.\\
\end{aligned}
\end{equation*}
For $A$ type potential, $I$ is the whole set of mesons. For $D$ type potential, $I$ denotes the combination $u^av^b$ such
that $ab$ is even.

\subsubsection{\textbf{A} gauge group with a pair of symmetric (and its conjugate) $(X, \tilde{X})$}
\begin{figure}[H]
	\centering
       \begin{tikzcd}
       |[draw,rectangle]| N_f\ar[r,"Q"]&    
       |[draw,circle]| N_c \ar[loop,out =290,in=250,looseness=5,"\tilde{X}"]\ar[loop,out =110,in=70,looseness=5,"X"] \ar[r,"\tilde{Q}"]&
        |[draw,rectangle]| N_f 
       \end{tikzcd}
       \begin{tikzcd}
		\stackrel{}{\longrightarrow}    
	\end{tikzcd}       
    \begin{tikzcd}
        |[draw,rectangle]| N_f\ar[loop,out =110,in=70,looseness=5,""]\ar[rr,bend right=80,""]&    
       |[draw,circle]| N_c^{'} \ar[loop,out =290,in=250,looseness=5,"\tilde{X}^*"]\ar[loop,out =110,in=70,looseness=5,"X^*"]\ar[l,"q"'] &
        |[draw,rectangle]| N_f \ar[loop,out =110,in=70,looseness=5,""]\ar[l,"\tilde{q}"']
       \end{tikzcd}
       \caption{\textbf{A} gauge group with a pair of symmetric and its conjugate matters.}
        \label{Agaugesym}
\end{figure}
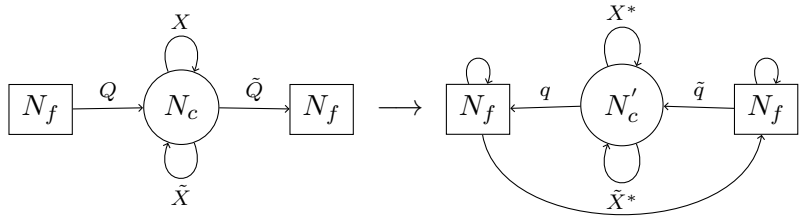
First, we have the anomaly free condition for 
$U(1)_R$ symmetry:
\begin{equation*}
\begin{aligned}
 & N_c+(\frac{N_c}{2}+1)\sum_u (R_u-1)+(R_c-1)N_f=0,\\
  &N_c^{'}+(\frac{N_c^{'}}{2}+1)\sum_u (R_u-1)+(R_{c^*}-1)N_f=0\,.\\ 
  \end{aligned}
\end{equation*}
We have 
\begin{equation*}
\begin{aligned}
&N_c^{'}=aN_f-N_c-\frac{4x}{x+2},\\
&R_c=1-\frac{N_c(1+x/2)+x}{N_f}\,.
\end{aligned}
\end{equation*}
Here $x=\sum_u(R_u-1)$ and $a=\frac{\Delta+2}{x+2}$. $R_c$ and $R_{c^*}$ are still paired as $R_c+R_{c^*}=1-{\Delta\over 2}$, and $\Delta$ is the pairing constant of the (undressed) mesons. 

If there is a single pair of symmetric matter, with the superpotential $\Tr(X\tilde{X})^{k+1}$, then
the dressed mesons are $(M_j)^{f\dot g}\equiv Q^{f}(\tilde{X}X)^{j}\tilde{Q}^{\dot{g}},~j=0,\dots ,k$, $(P_{r}) ^{fg}\equiv Q^{f}( \tilde{X} X) ^{r}\tilde{X} Q^{g}$ and $( \tilde{P} _{r}) ^{\dot{f} \dot{g} }\equiv \tilde{Q} ^{\dot{f} }X( \tilde{X} X) ^{r}\tilde{Q} ^{\dot{g} }, r= 0,\dots ,k- 1$ where $M_j$'s are adjoint and $P_r,\tilde P_r$'s are symmetric tensors of the flavor group. There are $k+1$ adjoint mesons, and $2k$ symmetric mesons. The transformation property is that of the flavor group.

For Figure \ref{Agaugesym} we have $x=2R_u-2,~\Delta=2kR_u=(2+x)k$ and thus $a=2k+1$ \footnote{In this case, the number $a$ appeared in the dual gauge group rank is not equal to the number of mesons in the spectrum.}. So the anomalies are
\begin{equation*}
\begin{aligned}
\Tr R&=N_c^2+\frac{1}{2}(N_c^2+N_c)\sum_u(R_u-1)+2N_f N_c(R_c-1),\\ 
\Tr R_{\text{dual}}&=N_c^{'2}+\frac{1}{2}(N_c^{'2}+N_c^{'})\sum_u(R_u-1)+2N_f N_c^{'}(R_{c^*}-1)\\
&+N_f^2\sum_{\text{adjoint mesons}}(2R_c+R_j-1)+\frac{N_f^2+N_f}{2}\sum_{\text{symmetric mesons}}(2R_c+R_j-1)\\
&=N_c^{'2}+\frac{1}{2}(N_c^{'2}+N_c^{'})\sum_u(R_u-1)+2N_f N_c^{'}(R_{c^*}-1)\\
&+N_f^2 (k+1)  (2R_c-1+\Delta/2)  +\frac{N_f^2+N_f}{2}2k(2R_c-1+\Delta/2)\,.
\end{aligned}
\end{equation*}
Then one can finally check that $\Tr R_{\text{dual}}=\Tr R$.

\subsubsection{\textbf{A} gauge group with  pairs of anti-symmetric (and its conjugate) $(X, \tilde{X})$}

First, we have the anomaly free condition for 
$U(1)_R$ symmetry:
\begin{equation*}
\begin{aligned}
  &N_c+(\frac{N_c}{2}-1)\sum_u (R_u-1)+(R_c-1)N_f=0,\\
  &N_c^{'}+(\frac{N_c^{'}}{2}-1)\sum_u (R_u-1)+(R_{c^*}-1)N_f=0\,.\\ 
  \end{aligned}
\end{equation*}
We have 
\begin{equation*}
\begin{aligned}
&N_c^{'}=aN_f-N_c+\frac{4x}{x+2},\\
&R_c=1-\frac{N_c(1+x/2)-x}{N_f}\,.
\end{aligned}
\end{equation*}
Here $x=\sum_u(R_u-1)$ and $a=\frac{\Delta+2}{x+2}$. $R_c$ and $R_{c^*}$ are still paired as $R_c+R_{c^*}=1-{\Delta\over 2}$, and $\Delta$ is the pairing constant of the (undressed) mesons. 

If there is a single pair of symmetric matter, with the superpotential $\Tr(X\tilde{X})^{k+1}$, then
the dressed mesons are $(M_j)^{f\dot g}\equiv Q^{f}(\tilde{X}X)^{j}\tilde{Q}^{\dot{g}},~j=0,\dots ,k$, $(P_{r}) ^{fg}\equiv Q^{f}( \tilde{X} X) ^{r}\tilde{X} Q^{g}$ and $( \tilde{P} _{r}) ^{\dot{f} \dot{g} }\equiv \tilde{Q} ^{\dot{f} }X( \tilde{X} X) ^{r}\tilde{Q} ^{\dot{g} }, r= 0,\dots ,k- 1$ where $M_j$'s are adjoint and $P_r,\tilde P_r$'s are antisymmetric tensors of the flavor group.

For Figure \ref{Agaugesym} we have $x=2R_u-2,~\Delta=2kR_u=(2+x)k$ and thus $a=2k+1$. So the anomalies are
\begin{equation*}
\begin{aligned}
\Tr R&=N_c^2+\frac{1}{2}(N_c^2-N_c)\sum_u(R_u-1)+2N_f N_c(R_c-1),\\ 
\Tr R_{\text{dual}}&=N_c^{'2}+\frac{1}{2}(N_c^{'2}-N_c^{'})\sum_u(R_u-1)+2N_f N_c^{'}(R_{c^*}-1)\\
&+N_f^2\sum_{\text{adjoint mesons}}(2R_c+R_j-1)+\frac{N_f^2-N_f}{2}\sum_{\text{anti-symmetric mesons}}(2R_c+R_j-1)\\
&=N_c^{'2}+\frac{1}{2}(N_c^{'2}-N_c^{'})\sum_u(R_u-1)+2N_f N_c^{'}(R_{c^*}-1)\\
&+N_f^2 (k+1)  (2R_c-1+\Delta/2)  +\frac{N_f^2-N_f}{2}2k(2R_c-1+\Delta/2)\,.
\end{aligned}
\end{equation*}
Then one can finally check that $\Tr R_{\text{dual}}=\Tr R$.

\section{Formula for superconformal index}
\label{index}
The formula relevant for the computation of the large $N_c, N_f$ index:

\begin{align*}
    &g_E=\bar g_E=\frac{t^{R_c}-t^{2-R_c}}{(1-tx)(1-tx^{-1})},\\
    &g_M=\bar g_M=\frac{t^{R_{c^*}}-t^{2-R_{c^*}}}{(1-tx)(1-tx^{-1})},\\
    &f=\frac{2t^2-t(x+x^{-1})+\sum_{u=1}^{N_A}(t^{R_u}-t^{2-R_u})}{(1-tx)(1-tx^{-1})},\\
    &h(M_{I})=\frac{t^{2R_c+R_j}-t^{2-(2R_c+R_j)}}{(1-tx)(1-tx^{-1})}\,.
\end{align*}
Here $R_c$ is the $R$ charge for the fundamental fields $c$, and $R_{c^*}$ is the 
$R$ charge for the dual fundamental fields $c^*$. $R_u$'s are the $R$ charges for 
the adjoint fields and their values are the same for electric and magnetic theory. Finally, 
$M_I$ denotes the dressed meson whose $R$ charge is $2R_c+R_j$, with $R_j$ the R charge for 
the meson $u_I$ formed by adjoint chiral field.

\bibliographystyle{JHEP}
\bibliography{ref1}

\end{document}